\documentclass[sigconf]{acmart}

\usepackage{expl3}
\expandafter\def\csname ver@l3regex.sty\endcsname{} 
\usepackage{amsmath}

\usepackage{amssymb,amsfonts}
\usepackage{algorithmic}
\usepackage{graphicx}
\usepackage{textcomp}
\usepackage{xcolor}
\def\BibTeX{{\rm B\kern-.05em{\sc i\kern-.025em b}\kern-.08em
    T\kern-.1667em\lower.7ex\hbox{E}\kern-.125emX}}
\usepackage[utf8]{inputenc}
\usepackage{booktabs} 
\usepackage{graphicx}
\usepackage{subfigure}
\usepackage{units}
\usepackage{blindtext}
\usepackage{multirow}
\usepackage{xcolor,xspace}
\usepackage{amsmath}
\usepackage{paralist}
\usepackage{mdframed}
\usepackage{pict2e}
\usepackage{mdframed}
\usepackage{listings}
\usepackage{color}
\usepackage{threeparttable}
\usepackage{array}
\usepackage{booktabs}
\usepackage{pifont}
\usepackage{balance}
\usepackage{url}

\usepackage{lipsum}

\usepackage[ruled]{algorithm2e}
\usepackage{algorithmic}
\LinesNumbered

\usepackage{url}

\usepackage[
    n,
    advantage,
    operators,
    sets,
    adversary,
    landau,
    probability,
    notions,
    logic,
    ff,
    mm,
    primitives,
    events,
    complexity,
    asymptotics,
    keys]{cryptocode}
                    
\usetikzlibrary{arrows, patterns, automata}

\newcommand{\ignore}[1]{}

\newcommand{\acro}{{{\it $RATA$}}\xspace}
\newcommand{\acroa}{{{\it {$RATA_{A}$}}}\xspace}
\newcommand{\acrob}{{{\it {$RATA_{B}$}}}\xspace}

\newcommand{\vrased}{{{\sf\it VRASED}}\xspace}

\newcommand{\styleAlgorithm}[1] {\textbf{\texttt{#1}}\xspace}

\newcommand{\OParan}{\styleAlgorithm{(}}
\newcommand{\CParan}{\styleAlgorithm{)}}

\newcommand{\dev}{{\ensuremath{\sf{\mathcal Prv}}}\xspace}
\newcommand{\prv}{{\ensuremath{\sf{\mathcal Prv}}}\xspace}
\newcommand{\vrf}{{\ensuremath{\sf{\mathcal Vrf}}}\xspace}

\newcommand{\RA}{{\ensuremath{\sf{\mathcal RA}}}\xspace}
\newcommand{\sadv}{{\ensuremath{\sf{\mathcal Adv}}}\xspace}
\newcommand{\chal}{{\ensuremath{\sf{\mathcal Chal}}}\xspace}
\newcommand{\auth}{{\ensuremath{\sf{\mathcal Auth}}}\xspace}

\newcommand{\attkey}{\ensuremath{\mathcal K}\xspace}

\renewcommand\adv{\ensuremath{\sf{\mathcal Adv}}\xspace}

\newcommand{\request}{\ensuremath{\mathsf{\bf Request}}\xspace}

\newcommand{\attest}{\ensuremath{\mathsf{\bf Attest}}\xspace}

\newcommand{\vrfy}{\ensuremath{\mathsf{\bf Verify}}\xspace}

\newcommand{\utoken}{{\ensuremath{\mathsf{H}}}\xspace}
\newcommand{\setup}{\ensuremath{\mathsf{\bf \underline{Setup}}}\xspace}
\newcommand{\challenge}{\ensuremath{\mathsf{\bf \underline{Challenge}}}\xspace}
\newcommand{\response}{\ensuremath{\mathsf{\bf \underline{Response}}}\xspace}

\mathchardef\mhyphen="2D

\newcommand{\toctou}{{\small TOCTOU}\xspace}
\newcommand{\hw}{\texttt{\small HW-Mod}\xspace}
\newcommand{\sw}{\texttt{\small SW-Att}\xspace}
\newcommand{\swtiny}{\texttt{\tiny SW-Att}\xspace}
\newcommand{\rom}{\texttt{ROM}\xspace}

\newcommand{\hmac}{HMAC\xspace}

\newcommand{\dmaaddr}{\ensuremath{DMA_{addr}}\xspace}
\newcommand{\dmaen}{\ensuremath{DMA_{en}}\xspace}

\newtheorem{construction}{Construction}

\setcopyright{none} 
\acmConference[To appear: ACM CCS]{}
\acmYear{2021}

\begin{document}
%
\title{On the TOCTOU Problem in Remote Attestation}


\author{Ivan De Oliveira Nunes}
\affiliation{%
   \institution{UC Irvine}}

\author{Sashidhar Jakkamsetti}
\affiliation{%
   \institution{UC Irvine}}

\author{Norrathep Rattanavipanon}
\affiliation{%
   \institution{Prince of Songkla University, Phuket Campus}}

\author{Gene Tsudik}
\affiliation{%
   \institution{UC Irvine}}   
   
\begin{abstract}
Much attention has been devoted to verifying software integrity of remote embedded (IoT) devices. Many techniques, with different 
assumptions and security guarantees, have been proposed under the common umbrella of so-called {\it Remote Attestation} (\RA).
Aside from executable's integrity verification, \RA serves as a foundation for many security services, such as proofs of memory 
erasure, system reset, software update, and verification of runtime properties. 
Prior \RA techniques verify the remote device's binary at the time when \RA functionality is executed, thus providing no information 
about the device's binary before current \RA execution or between consecutive \RA executions. This implies that presence of transient
malware (in the form of modified binary) may be undetected. In other words, if transient malware infects a device (by modifying its binary),
performs its nefarious tasks, and erases itself before the next attestation, its temporary presence {\bf will not be detected}. This 
important problem, called Time-Of-Check-Time-Of-Use (\toctou), is well-known in the research literature and remains unaddressed 
in the context of hybrid \RA. 

In this work, we propose \underline{R}emote \underline{A}ttestation with \underline{T}OCTOU \underline{A}voidance (\acro): a 
provably secure approach to address the \RA~\toctou problem. With \acro, even malware that erases itself before execution 
of the next \RA, can not hide its ephemeral presence. \acro targets hybrid \RA architectures, which are aimed at low-end 
embedded devices. We present two alternative techniques -- \acroa and \acrob~-- suitable
for devices with and without real-time clocks, respectively. Each is shown to be secure and accompanied by a publicly available and 
formally verified implementation. Our evaluation demonstrates low hardware overhead of both techniques. Compared with current 
hybrid \RA architectures -- that offer no \toctou protection -- \acro incurs no extra runtime overhead. In fact, it substantially 
reduces the time complexity of \RA computations: from linear to constant time.
\end{abstract}

\maketitle
\section{Introduction}\label{sec:intro}
In the last two decades, our society is gradually becoming surrounded by, and dependent upon, 
a multitude of small and specialized computing devices that perform a wide range of functions in many 
aspects of everyday life. They are often referred to as embedded, ``smart", CPS or IoT devices, and they 
vary  widely in terms of computing abilities. However, regardless of their purposes and resources, these devices 
have become popular targets for malicious exploits and malware.

\begin{figure}[ht]
\center
\resizebox{0.7\columnwidth}{!}{
	\centering\fbox{
	\begin{tikzpicture}[node distance=1.5cm, >=stealth]
	\coordinate (BL)	at (0, 0);		\coordinate[above of = BL]	 (TL);
	\coordinate (Btvs)	at (1, 0);		\coordinate[above of = Btvs] (Ttvs);
	\coordinate (Btdr)	at (1.75, 0);	\coordinate[above of = Btdr] (Ttdr);
	\coordinate (Btcs)	at (2.25, 0);		\coordinate[above of = Btcs] (Ttcs);
	\coordinate (Btce)	at (6.75, 0);	\coordinate[above of = Btce] (Ttce);
	\coordinate (Btds)	at (7.25, 0);	\coordinate[above of = Btds] (Ttds);
	\coordinate (Btvr)	at (8, 0);		\coordinate[above of = Btvr] (Ttvr);
	\coordinate (BR)	at (9, 0);		\coordinate[above of = BR]	 (TR);
	\coordinate (att) at ($(Btcs)!0.5!(Btce)$);

	\draw[line width = .4cm, color=gray!50]	(Btcs) -- (Btce);

	\node[left] at (BL) {{\ensuremath{\sf{\mathcal Prv}}}\xspace};
	\node[left] at (TL) {{\ensuremath{\sf{\mathcal Vrf}}}\xspace};
	\node[below = .15cm] at (att) {Authenticated integrity check};

	\draw[thick,->]	(BL) -- (BR);
	\draw[thick,->] (TL) -- (TR);
	\draw[thick, densely dashed]	(Btvs) -- (Ttvs)
									(Btdr) -- (Ttdr)
									(Btcs) -- (Ttcs)
									(Btce) -- (Ttce)
									(Btds) -- (Ttds)
									(Btvr) -- (Ttvr);
	\draw[thick, ->] (Ttvs) -- (Btdr) node [below=-.05cm, midway, sloped] {req};
	\draw[thick, ->] (Btds) -- (Ttvr) node [below=-.05cm, midway, sloped] {resp};
\end{tikzpicture}
	}
}
	\caption{\small Timeline of a typical \RA protocol}
	\label{fig:timeline}
\vspace{-3mm}
\end{figure}
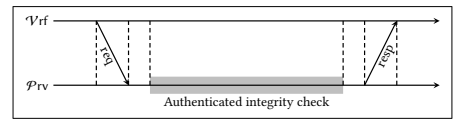

At the low-end of the spectrum, Micro-Controller Units (MCUs) are designed with strict constraints on monetary cost, 
physical size, and energy consumption.
These MCUs exist on the edge of more complex systems, typically interfacing digital and physical domains, and often 
implementing safety-critical sensing and/or actuation functions.
Two  prominent examples of such MCUs are: 
TI MSP430\footnote{\url{http://www.ti.com/microcontrollers/msp430-ultra-low-power-mcus/applications.html}} and 
Atmel ATMega AVR\footnote{\url{https://www.microchip.com/design-centers/8-bit/avr-mcus}}.
It is unrealistic to expect such devices, by themselves, to prevent malware infection via sophisticated security 
mechanisms (similar to those available on laptops, smartphones, or relatively higher-end/higher-power embedded systems\footnote{For 
instance, higher-end devices such as Raspberry Pi, Tessel, and similar, are usually 2-to-3 orders of magnitude more expensive 
in terms of monetary cost, physical size, and energy consumption than the MCUs targeted in this work.}).
In this landscape, {\bf Remote Attestation (\RA)} emerged as an inexpensive and effective means to detect unauthorized modifications 
to executables of remote low-end devices. Also, \RA serves as a foundation for other important security services, such as software 
updates~\cite{pure,ammar2020verify}, control-flow integrity verification~\cite{cflat, tinycfa, dessouky2018litehax},
and proofs of remote software execution~\cite{apex}. Generally speaking, \RA allows a trusted 
entity, called a Verifier (\vrf), to ascertain memory integrity of an untrusted remote device, called Prover (\prv). 
As shown in Figure~\ref{fig:timeline}, \RA is typically realized as a (deceptively) simple challenge-response protocol:
\begin{compactenum}
	\item \vrf sends an attestation request with a challenge (\chal) to \dev.
	This request might also contain a token derived from a secret that allows \dev to authenticate \vrf.
	\item \dev receives the request and computes a \chal-based {\em authenticated integrity check} over a 
		pre-defined memory region. In low-end embedded systems (and in the context of this paper) this region 
		corresponds to the entire executable memory, i.e., program memory. See Section~\ref{sec:MCU_assumptions}.
	\item \dev returns the result to \vrf.
	\item \vrf receives the result and checks whether it corresponds to a valid memory state.
\end{compactenum}

The {\em authenticated integrity check} requires some type of integrity-ensuring function, typically implemented as a 
Message Authentication Code (MAC), computed over \dev's attested memory region. Computing a MAC requires \dev to have a 
unique secret key, denoted by \attkey --
a symmetric key shared with \vrf, or a private key for which the corresponding public key is known to \vrf.
\attkey must reside in secure storage, and be {\bf inaccessible} to any software running on \dev, except for privileged and immutable
attestation code.
Since the usual \RA threat model assumes a fully compromised software state on \dev, secure storage implies some 
level of hardware support. Hybrid RA (based on hardware/software co-design)~\cite{smart,trustlite,hydra,vrasedp} is an 
approach particularly suitable for low-end embedded devices. In hybrid \RA designs, the integrity-ensuring function 
is implemented in software, while hardware controls execution of this software, detecting any violations that might cause 
unexpected behavior or \attkey leakage. In a nutshell, hybrid \RA provides the same security guarantees as (more expensive) 
hardware-based \RA approaches (e.g., those based on a TPM~\cite{tpm} or other standalone hardware modules), 
while minimizing modifications to the underlying 
hardware platform. We overview a concrete hybrid \RA architecture in Section~\ref{sec:background}.

Despite much progress, current hybrid \RA architectures share a common limitation: they measure the state of \prv's executables 
at the time when \RA is executed by \prv. They provide no information about \prv's executables \textbf{before} \RA measurement or its state in
\textbf{between} two consecutive \RA measurements. We refer to this problem as \emph{Time-Of-Check Time-Of-Use} or  \toctou.
While variants of this problem have been discussed before~\cite{erasmus,LISA,FNR+14,toctou,survey}, it remains 
unsolved in the context of hybrid \RA.

We emphasize that the \RA--\toctou problem (as formulated in this paper) should not be confused or conflated with the problem of 
ensuring temporal consistency between attestation and  execution of a binary, which is solved by runtime attestation approaches, e.g., 
\cite{apex,cflat,tinycfa,dessouky2018litehax}. Nonetheless, \RA of binaries (i.e., static \RA) is still relevant in that context because 
most runtime attestation techniques for low-end devices rely on static \RA as a building block (one exception is~\cite{geden2019hardware},
which, instead, assumes that binaries never change~\footnote{In many settings involving low-end MCUs this assumption is unrealistic. See Section~\ref{sec:scope} for details.}). In fact, as we discuss in Section~\ref{sec:discussion}, an \RA architecture secure 
against \toctou makes runtime attestation techniques that rely on static \RA substantially more efficient.

In practice, the \toctou problem leaves devices vulnerable to transient malware which erases itself (its executable) after completing its tasks.
This is harmful in settings where numerous MCUs report measurements collected over extended periods.
For example, consider several MCU-based sensors that measure energy consumption in a smart city, where large scale 
erroneous measurement may lead to power outages.
If regular \RA schemes that are not secure against \toctou are used in this case (e.g., by performing \RA once a day, or 
once per billing cycle), security can be subverted by {\bf (i)} changing the sensor software to spoof measurements during 
regular usage, and {\bf (ii)} reprogramming the sensor back with the expected executable
immediately before the scheduled \RA computation.  In particular, since the \RA request must be received through an 
\emph{untrusted} communication channel, malware may simply erase itself upon detecting an incoming attestation request, 
even if \RA schedule is not known \textit{a priori}. As noted earlier, in settings where detection of runtime violations 
(e.g., code-reuse and data corruption attacks) is also desirable (e.g., MCU code is written with memory-unsafe languages), 
\toctou-Security of the underlying static \RA makes the overall runtime attestation more efficient (see Section~\ref{sec:discussion}).

Our approach to solve the \toctou problem is rooted in the observation that current hybrid \RA 
techniques use trusted hardware only to detect security violations that compromise execution of \RA software itself and 
take action (e.g., by resetting the device) if such a violation is detected. Whereas, \acro's main new feature is the use of a minimal 
(formally verified) hardware component to additionally provide historical context about the state of \prv's program memory.
This is achieved via secure logging of the latest timing of program memory modifications in a protected memory 
region that is also covered by \RA's integrity-ensuring function. This enables \vrf to later check authenticity and integrity of 
\prv's memory modifications. This new feature is integrated 
seamlessly into the underlying \RA architecture and the composition is shown to be secure. We believe this results in the 
following contributions: 

	\noindent-- \textbf{RA~\toctou-Security Formulation:} We motivate and formalize \toctou in the context of \RA. We define 
	\RA~\toctou-Security using a security game (see Definition~\ref{def:toctou_sec}) and discuss why current \RA techniques 
	based on consecutive self-measurements do not satisfy this definition. We believe our work to be the first formal treatment 
	of this matter. Furthermore, we evaluate practicality of RA techniques based on consecutive self-measurements 
	and argue that using them to obtain \toctou-Security incurs extremely high runtime overhead, possibly starving 
	benign applications on \prv.

	\noindent-- \textbf{RATA Design, Implementation \& Verification:} We propose two techniques: \acroa and \acrob.
	The former assumes that \prv has a secure read-only Real-Time Clock (RTC) synchronized with \vrf. 
	Since this assumption is unrealistic for many low-end \prv-s, we construct \acrob\ which trades off the need for a secure clock 
	for the need to authenticate \vrf's attestation requests; this feature is already included in several hybrid \RA architectures.   
	
	We show that both techniques satisfy the formal definition of \toctou-Security, assuming that their implementations 
	adhere to a set of formal specifications, stated in Linear Temporal Logic (LTL). Finally, the implementation itself is formally 
	verified to adhere to these LTL specifications, yielding security at both design and implementation levels. 
	Our implementation is publicly available at~\cite{repo}. It is realized on a real-world low-end MCU -- TI MSP430 -- and is 
	deployed using commodity FPGAs. Experimental results show low hardware overhead, 
	affordable even for cost-sensitive low-end devices.

	\noindent-- \textbf{RATA Enhancements to \RA and Related Services:} We discuss the implications of \acro on \RA and related 
	services beyond \toctou-Security. In particular, we show that \acro can, in most cases, lower \RA computational complexity 
	from linear (in terms of attested memory size) to constant time, resulting in significant savings. 
	We also discuss \acro's benefits for specialized \RA applications: (i) real-time systems; (ii) run-time integrity/control-flow 
	attestation; and (iii) collective \RA, where a multitude of provers need to be attested simultaneously.

\section{Scope}\label{sec:scope}
This section delineates the scope of this paper in terms of targeted devices and desired security properties.

\noindent\textbf{Low-End Devices:}
This work focuses on CPS/IoT sensors and actuators (or hybrids thereof) with low computing power.
These are some of the smallest and weakest devices based on low-power single-core MCUs with only a few KBytes 
of program and data memory. Two prominent examples are: Atmel AVR ATmega and TI MSP430:
$8$- and $16$-bit CPUs, typically running at $1$-$16$MHz clock frequencies, with $\approx64$ KBytes of addressable memory.
SRAM is used as data memory with the size normally ranging between $4$ and $16$KBytes, while the rest of address space is 
available for program memory.  Such devices usually run software atop ``bare metal'' and execute instructions in place 
(physically from program memory) and have no memory management unit (MMU) to support virtual memory.

Our implementation is based on MSP430. This choice is due to public availability of a well-maintained open-source MSP430 
hardware design from Open Cores \cite{openmsp430}. Nevertheless, our machine model and the entire methodology developed in 
this paper are applicable to other low-end MCUs in the same class, such as Atmel AVR ATmega. \acro's main implementation is composed with \vrased, a publicly available verified hybrid \RA architecture \cite{vrasedp}, which allows us to demonstrate security.
Despite our specific implementation choices, we believe that \acro's concepts are also applicable to other \RA 
architectures.
To support this claim, in Appendix~\ref{apdx:sancus} we report on \acro's implementation atop SANCUS \cite{Sancus17}: a hardware-based \RA architecture targeting low-end devices; see Section~\ref{sec:rw} for an overview of types of \RA architectures.

\vspace{2mm}
\noindent\textbf{Detection, Prevention \& Memory Immutability:}
As a detection-oriented security service, \RA does not prevent future binary modifications. Therefore, the term \toctou should be 
considered in retrospective. In particular, techniques presented in this paper allow \vrf to understand \underline{``since when''}  
\prv's memory remained the same as reported in the present \RA result. 

While malware infections can be trivially prevented by making all executable memory read-only (e.g., storing code in ROM), such a drastic approach 
would sacrifice reconfigurability: it would make legitimate software updates impossible and would essentially transform the MCU into an Application-Specific Integrated Circuit (ASIC).
However, reconfigurability is one of the most important MCU features, perhaps even its entire ``raison d'\^etre".

A less drastic approach is to prevent program memory modifications that occur at runtime. This approach is vulnerable to modifications by adversaries with physical access to re-program \prv directly. More importantly, (even if attacks that require physical access are out of scope) it makes remote updates impossible, requiring physical access whenever a device's binary needs to be updated.
Since these devices are often remote or physically inaccessible (inside a larger systems, e.g., a vehicle) low-end MCUs (including aforementioned MSP430 and ATMega) typically do not prevent modifications to program memory. Our detection-based approach conforms with that necessity, allowing changes to binaries and reporting them to \vrf: even if they happen in between subsequent attestations. In turn, \vrf is able to detect binary changes, and tell apart illegal modifications from expected ones.

\section{Background \& Definitions}\label{sec:prelim}
\subsection{Device Model \& MCU Assumptions}\label{sec:mcu_model}
\label{sec:MCU_assumptions}
Below, we overview MCU assumptions relevant to \acro. They reflect the behavior of the class of low-end embedded 
systems discussed in Section~\ref{sec:scope} and are in accordance with previous work on securing low-end
MCUs~\cite{smart,NAD+13,vrasedp,apex,pure}. In particular, we assume that the MCU hardware correctly implements 
its specifications, as follows:\\
\noindent {\bf A1 -- \emph{Program Counter (PC):}} $PC$ always contains the address of the instruction being executed 
in a given CPU cycle.

\noindent {\bf A2 -- \emph{Memory Address:}} Whenever memory is read or written, a data-address signal ($D_{addr}$) 
contains the address of the corresponding memory location. For a read access, a data read-enable bit ($R_{en}$) 
must be set, while, for a write access, a data write-enable bit ($W_{en}$) must be set.
 
\noindent {{\bf A3 -- \emph{DMA:}} Whenever the Direct Memory Access (DMA) controller attempts to access the main system memory, a 
DMA-address signal (\dmaaddr) reflects the address of the memory location being accessed and the DMA-enable bit 
(\dmaen) is set. DMA can not access memory without setting \dmaen.
 
\noindent {\bf A4 -- \emph{MCU Reset:}} At the end of a successful reset routine, all registers (including $PC$) are set to zero 
before resuming normal software execution flow. Resets are handled by the MCU in hardware. Thus, the reset handling 
routine can not be modified. When a reset happens, the corresponding $reset$ signal is set. The same signal is also 
set when the MCU initializes for the first time. 
 
\noindent {\bf A5 -- \emph{No Data Execution:}} Instructions \textbf{\underline{must}} reside (physically) in program memory 
(PMEM) in order to execute. They are not loaded to DMEM to execute. Data execution is impossible in most low-end devices, 
including OpenMSP430 used in our prototype. For example, in Harvard-based low-end devices (e.g., AVR Atmega), there is no 
hardware support to fetch/execute instructions from data memory (DMEM). In other low-end devices that do not prevent 
data execution by default, this is typically enforced by the underlying hybrid RA architecture. Therefore, even if malware 
resides in DMEM, it must be copied to, and thus reside in, PMEM before executing.
\subsection{\RA Definitions, Architectures \& Adversarial Model}\label{sec:background}
As discussed in Section~\ref{sec:intro}, \RA is typically realized as a challenge-response protocol between 
\vrf (challenger) and \prv, a potentially compromised remote low-end device. This notion is captured by a generic 
syntax for \RA protocols in Definition~\ref{def:ra}.

\begin{figure}[!hbtp]
\begin{mdframed}
\footnotesize
\begin{definition}[syntax]\label{def:ra}
	\RA is a tuple ($\request$, $\attest$, $\vrfy$) of algorithms:
	\begin{compactitem}~
		\item $\request^{\vrf \rightarrow \dev}\OParan \cdots \CParan$: algorithm initiated by \vrf to request a measurement of \dev memory 
		range $AR$ (attested range). As part of \request, \vrf sends a challenge \chal to \dev.
		\item $\attest^{\dev \rightarrow \vrf}\OParan \chal, \cdots \CParan:$ algorithm executed by \dev upon receiving \chal from \vrf. 
		Computes an authenticated integrity-ensuring function over $AR$ content. It produces attestation token \utoken, which is returned 
		to \vrf, possibly accompanied by auxiliary information to be used by the \vrfy algorithm (see below).
		\item $\vrfy^{\vrf}\OParan \utoken, \chal, M, \cdots \CParan:$ algorithm executed by \vrf upon receiving \utoken from \prv. It verifies 
		whether \prv's current $AR$ content corresponds to some expected value $M$ (or one of a set of expected values). 
		\vrfy outputs: $1$ if \utoken is valid, and $0$ otherwise.
	\end{compactitem}
	\vspace{1mm}
	\textbf{Note:} In the parameter list, ($\cdots$) denotes that additional parameters might be included, depending on the specific \RA construction.
\end{definition}
\end{mdframed}
\vspace{-2mm}
\end{figure}

Definition~\ref{def:ra} specifies \RA as a tuple (\request, \attest,\vrfy). \request is computed by \vrf to produce challenge \chal and 
send it to \dev. \attest is performed by \prv by using \chal to compute an authenticated integrity-ensuring function (e.g., MAC) 
over attested memory range (denoted by $AR$) and producing \utoken, which is sent back to 
\vrf for verification. For example, if \attest is implemented using a MAC, \utoken is computed as:
%
\begin{equation}
 \utoken = MAC_{\attkey}(\chal||AR)
\end{equation}
where $||$ denotes concatenation and \attkey is a symmetric key shared by \prv and \vrf.
Upon receiving \utoken, \vrf executes algorithm \vrfy by checking if \utoken corresponds to the $MAC$ of some expected value $M$.

Although techniques discussed in this paper are not tied to a specific \RA architecture, we chose to compose 
\acro with \vrased~\cite{vrasedp}. Our choice is motivated by \vrased's formal security definitions, which allow 
reasoning about \acro's secure composition with the underlying \RA architecture; see Theorems~\ref{th:toctou_sec} 
and~\ref{th:toctou_sec_b}. We overview \vrased\ next.

\vrased is a formally verified hybrid RA architecture, based on a hardware/software co-design.
It is built as a set of sub-modules, each guaranteeing a specific set of sub-properties.
Every sub-module (hardware or software) is individually verified. Finally, composition of all sub-modules 
is proved to satisfy formal definitions of \RA soundness and security. Informally, \RA soundness guarantees that an integrity-ensuring 
function (\hmac in \vrased's case) is correctly computed over attested memory range ($AR$). It also guarantees that $AR$ 
can not be modified after the start of \RA computation, thus enforcing temporal consistency and protecting against 
``hide-and-seek'' attacks during \RA computation~\cite{carpent2018temporal}. \RA security ensures that \RA execution 
generates an unforgeable authenticated memory measurement and that \attkey used in computing this measurement 
is not leaked before, during, or after, attestation.

To achieve its aforementioned goals, \vrased's software part (\sw) resides in Read-Only Memory (\rom) and 
relies on a formally verified \hmac implementation from the HACL* cryptographic library~\cite{hacl}.
A typical \sw execution proceeds as follows:
\begin{compactenum}
	\item Read challenge \chal from a fixed memory region denoted by $MR$. 
	\item Use a Key Derivation Function (KDF) to derive a one-time key from \chal and the attestation master key \attkey: 
		$KDF(\attkey, MR)$ (where $MR=\chal$).
	\item \attest's implementation (\sw) generates attestation token \utoken by computing an HMAC over an 
		attested memory region $AR$ using the newly derived key:\\
		\centerline{$\utoken = HMAC(KDF(\attkey, MR), AR)$}
	\item Overwrite $MR$ with the result \utoken and return execution to unprivileged software, i.e, the normal application(s).
\end{compactenum}
\noindent \vrased's Hardware (\hw) monitors $7$ distinct MCU signals:
\begin{compactitem}
\item $PC$: Current Program Counter value;
\item $R_{en}$: Signal that indicates if the MCU is reading from memory (1-bit);
\item $W_{en}$: Signal that indicates if the MCU is writing to memory (1-bit);
\item $D_{addr}$: Address for an MCU memory access;
\item \dmaen: Signal that indicates if  Direct Memory Access (DMA) is currently accessing memory (1-bit);
\item \dmaaddr: Memory address being accessed by DMA.
\item $irq$: Signal that indicates if an interrupt is happening (1-bit);
\end{compactitem}
These signals determine a one-bit $reset$ signal output, that, when set to $1$, triggers an immediate system-wide 
MCU reset, i.e., before executing the next instruction. The $reset$ output is triggered when \vrased's hardware detects 
any violation of security properties. \vrased's hardware is described in Register Transfer Level (RTL) using Finite State 
Machines (FSMs). Then, NuSMV Model Checker~\cite{nusmv} is used to automatically prove that FSMs achieve 
claimed security sub-properties. Finally, the proof that the conjunction of hardware and software sub-properties implies 
end-to-end soundness and security is done using an LTL theorem prover.
\begin{figure}[!ht]
\begin{mdframed}
\footnotesize
\begin{definition}{\vrased's Security Game (Adapted from~\cite{vrasedp})}\label{def:vrased_sec}~\\
\textbf{Notation:}\\
- $l$ is the security parameter and $|\attkey| = |\chal| = |MR| = l$\\
- $AR(t)$ denotes the content of $AR$ at time $t$\\
\texttt{\RA-game:}
	\begin{compactenum}
	\item[]
	\item \texttt{\setup:} \adv\ is given oracle access to \attest (\sw) calls.
	\item \texttt{\challenge:} A challenge $\chal$ is generated by calling \request (Definition~\ref{def:ra}) and given to \adv.
	\item \texttt{\response:} \adv\ responds with a pair $(M, \sigma)$, where $\sigma$ is either forged by \adv, or is 
		the result of calling \attest (Definition~\ref{def:ra}), at some arbitrary time $t$.
	\item \adv\ wins iff $M \neq AR(t)$ and $\sigma = HMAC(KDF(\attkey, \chal), M)$.
	\end{compactenum}
\vspace{1mm}
\footnotesize{
\textbf{Note:} If, as a part of \attest, $AR$ attestation is preceded by a procedure to authenticate \vrf, 
$t$ defined in step 3 is the time immediately after successful authentication, 
when $AR$ attestation starts.}
\end{definition}
\end{mdframed}
\vspace{-2mm}
\end{figure}
More formally, \vrased end-to-end security proof guarantees that no probabilistic polynomial time (PPT) adversary can 
win the \RA \emph{security game} in Definition~\ref{def:vrased_sec} with non-negligible probability in the security parameter $l$, i.e.,
$Pr[\adv, \text{\RA-game}] \leq \negl[l]$.

\noindent\emph{\textbf{Remark 1:} While aforementioned guarantees ensure consistency of attested memory \underline{during} 
attestation computation, \vrased or any prior low-end \RA scheme is  \underline{not} \toctou-Secure, as 
modifications \underline{before} attestation remain undetected.}\\

\noindent\textbf{Adversarial Model.}
We consider a fairly strong adversary \adv\ that controls the entire software state of \dev, including both code and data.
\adv\ can modify any writable memory and read any memory (including secrets) that is not explicitly protected by 
trusted hardware.  Also, \adv\ has full access to all DMA controllers, if any are present on \dev. Recall that DMA 
allows direct access and memory 
modifications without going through the CPU.

Even though \adv\ may physically re-program \dev's software through wired connection to flash, invasive/tampering 
hardware attacks are out of scope of this paper: we assume that \adv\ can not: (1) alter hardware components, (2) 
modify code in \rom, (3) induce hardware faults, or (4) retrieve \dev secrets via physical side-channels.
Protection against physical hardware attacks is orthogonal to our goals and attainable via tamper-resistance 
techniques~\cite{ravi2004tamper}.

\subsection{Linear Temporal Logic (LTL)}\label{sec:prelim-fv}
Computer-aided formal verification typically involves three basic steps: {\bf First}, the system of interest (e.g., hardware, software, 
communication protocol) is described using a formal model,  e.g., a Finite State Machine (FSM). {\bf Second}, properties that 
the model should satisfy are formally specified. {\bf Third}, the system model is checked against formally specified properties to 
guarantee that it retains them. This can be achieved via either Theorem Proving or Model Checking.
In this work, we use the latter to verify the implementation of system modules.

In one instantiation of model checking, properties are specified as \textit{formulae} using Linear Temporal Logic (LTL) and system models 
are represented as FSMs. Hence, a system is represented by a triple $(S, S_0, T)$, where $S$ is a finite set of states,
$S_0 \subseteq S$ is the set of possible initial states, and $T \subseteq S \times S$ is the transition relation set -- it describes 
the set of states that can be reached in a single step from each state. The use of LTL to specify properties allows representation 
of expected system behavior over time.

In addition to propositional connectives, such as conjunction ($\land$), disjunction ($\lor$), negation ($\neg$), and implication 
($\rightarrow$), LTL includes temporal connectives, thus enabling sequential reasoning. In this paper, we are interested in 
the following temporal connectives:
\begin{compactitem}
	\item \textbf{X}$\phi$ -- ne\underline{X}t $\phi$: holds if $\phi$ is true at the next system state.
	\item \textbf{F}$\phi$ -- \underline{F}uture $\phi$: holds if there exists a future state where $\phi$ is true.
	\item \textbf{G}$\phi$ -- \underline{G}lobally $\phi$: holds if for all future states $\phi$ is true.
	\item $\phi$ \textbf{U} $\psi$ -- $\phi$ \underline{U}ntil $\psi$: holds if there is a future state where $\psi$ holds and
 	$\phi$ holds for all states prior to that.
	\item $\phi$ \textbf{W} $\psi$ -- $\phi$ \underline{W}eak until $\psi$: holds if, assuming a future state where $\psi$ 
		holds, $\phi$ holds for all states prior to that. If $\psi$ never becomes true, $\phi$ must hold forever. 
		More formally: $\phi \textbf{W} \psi \equiv (\phi \textbf{U} \psi) \lor \textbf{G}(\phi)$
\end{compactitem}

\section{\RA~TOCTOU}\label{sec:toctou}
This section defines the notion of \toctou-Security in the context of \RA.
We start by formalizing this notion using a security game.
Next, we consider the practicality of this problem and overview existing mechanisms, arguing that they do not 
achieve \toctou-Security (neither according to \toctou-Security definition, nor in practice) and incur high overhead.

\subsection{Notation}
We summarize our notation 
in Table~\ref{tab:notation}. We keep it mostly consistent with that in \vrased~\cite{vrasedp}, with a few additional 
elements to denote \acro-specific memory regions and signals. To simplify the notation, when the value of a given 
signal (e.g., $D_{addr}$) is within a certain range (e.g., $AR = [AR_{min}, AR_{max}]$), we write that $D_{addr} \in AR$, i.e.:
\begin{equation}
\footnotesize
 D_{addr} \in AR \quad \equiv \quad AR_{min} \leq D_{addr} \leq AR_{max}
\end{equation}
\sloppy
In conformance with axioms discussed in Section~\ref{sec:MCU_assumptions}, we use $Mod\_Mem(x)$ to denote a 
modification to memory address address $x$. Given our machine model, the following logical equivalence holds:
\begin{equation}
\footnotesize
Mod\_Mem(x) \equiv (W_{en} \land D_{addr}=x) \lor (DMA_{en} \land DMA_{addr}=x)
\end{equation}
this captures the fact that a memory modification can be caused by either the CPU (reflected in signals $W_{en}=1$ and 
$D_{addr}=x$) or by the DMA (signals $DMA_{en}=1$ and $DMA_{addr}=x$). We also use this notation to represent a 
modification to a location within a contiguous memory region $R$ as:
\begin{equation}
\footnotesize
Mod\_Mem(R) \equiv (W_{en} \land D_{addr}\in R) \lor (DMA_{en} \land DMA_{addr} \in R)
\end{equation}
%
%

\begin{table}[!htb]
    \caption{Notation}
    \vspace{-2mm}
    \begin{center}
         \fbox{\scriptsize
         \begin{tabular}{r p{6.3cm} }
              \multicolumn{2}{c}{}\\
              $PC$                                    &  Current Program Counter value                        \\
              $R_{en}$                                        &  Signal that indicates if the MCU is reading from memory (1-bit)              \\
              $W_{en}$                                        &  Signal that indicates if the MCU is writing to memory (1-bit)                \\
              $D_{addr}$                              &  Address for an MCU memory access             \\
              \dmaen                                  &  Signal that indicates if DMA is currently accessing memory (1-bit)                            \\
              \dmaaddr                                        &  Memory address being accessed by DMA, if any                                 \\
              $irq$                                           &  Signal that indicates if an interrupt is happening                           \\
              $CR$                                    &  Memory region where \swtiny is stored: $CR = [CR_{min}, CR_{max}]$   \\
              $MR$                            & (MAC Region) Memory region in which \swtiny computation result is written: $MR = [MR_{min}, MR_{max}]$.
              The same region is also used to pass the attestation challenge as input to \swtiny \\
              $AR$                                    & (Attested Region) Memory region to be attested.
              Corresponds to all executable memory (program memory) in the MCU: $AR = [AR_{min}, AR_{max}]$ \\
              $LMT$                                   & (Latest Modification Time) Memory region that stores a timestamp/challenge corresponding to the last $AR$ modification \\
              $CR_{Auth}$                             & The first instruction in \vrased's \sw that is executed after successful authentication of \vrf's request. \\
              $set_{LMT}$                    & (\acroa) A 1-bit signal overwrites $LMT$ with the current RTC time, when set to logical $1$.\\
              $UP_{LMT}$                    & (\acrob) A 1-bit signal overwrites $LMT$ with the content of $MR$ when set to logical $1$.\\
          \end{tabular}
         }
     \end{center}
    \label{tab:notation}
\end{table}

\subsection{TOCTOU-Security Definition}\label{sec:toctou_def}
Definition~\ref{def:toctou_sec} captures the notion of \toctou-Security. In it,
the game formalizes the threat model discussed in Section~\ref{sec:background}, 
where \adv controls \prv's entire software state, including the ability to invoke \attest at will. 
The game starts with the challenger (\vrf) choosing a time $t_0$. At a later time ($t_{att}$), 
\adv receives \chal and wins the game if it can produce $\utoken_{\sadv}$ that is accepted by \vrfy as a 
valid response for expected $AR$ value $M$, when, in fact, there was a time between $t_0$ and $t_{att}$ when $AR\neq M$.

\begin{figure}[!h]
\begin{mdframed}
\footnotesize
\begin{definition}\label{def:toctou_sec}~\\
\textbf{\ref{def:toctou_sec}.1 \RA-\toctou Security Game:}
Challenger plays the following game with \adv:
	\begin{enumerate}
	\item Challenger chooses time $t_0$.
	\item \adv is given full control over \dev software state 
	and oracle access to $\attest$ calls.
	\item At time $t_{att} > t_0$, \adv is presented with \chal.
	\item \adv wins if and only if it can produce $\utoken_{\adv}$, such that:
	\begin{equation}\label{eq:verif}
	\vrfy(\utoken_{\adv}, \chal, M, \cdots) = 1\
	\end{equation}
    \center{and}
	\begin{equation}\label{eq:exists}
    \exists_{t_0 \leq t_i \leq t_{att}}\{AR(t_i) \neq  M \}	 
	\end{equation}
	\end{enumerate}~
where $AR(t_i)$ denotes the content of $AR$ at time $t_i$.\\

\noindent \textbf{\ref{def:toctou_sec}.2 \RA-\toctou Security Definition:} 
An \RA scheme is considered \toctou-Secure if -- for all PPT adversaries \adv~-- there exists a negligible function 
$\negl[]$, such that:
	\begin{center}
		$Pr[\adv, \text{\RA-\toctou-game}] \leq \negl[l]$
	\end{center}
where $l$ is the security parameter.
\end{definition}
\end{mdframed}
\vspace{-2mm}
\end{figure}

This definition augments \RA security (Definition~\ref{def:vrased_sec}) to incorporate \toctou attacks, by additionally 
allowing \adv\ to win if it can produce the expected response and $AR$ was modified at any point after $t_0$, where $t_0$ is chosen by \vrf.
For example, if \vrf wants to know if $AR$ remained in a valid state for the past two hours, \vrf chooses $t_0$ as $t_0 = t_{att} - 2h$.
Note that this definition also captures security against transient attacks wherein \sadv changes modified memory back to its expected state 
and leaves the device, thus attempting to hide its ephemeral modification from the upcoming attestation request. 
This attack is undetectable by all \RA schemes that are not \toctou-Secure.

\noindent\emph{\textbf{Remark 2:} Recall that $AR$ corresponds to the executable part of \prv's memory, i.e., program memory. 
Since data memory is not executable (see Section~\ref{sec:MCU_assumptions}), changes to data memory are not taken into account 
by Definition~\ref{def:toctou_sec}. \acro's relation to runtime/data-memory attacks is discussed in Section~\ref{sec:app4}.}
\subsection{TOCTOU-Secure \RA vs. Consecutive Self-Measurements}
\label{sec:motivation}
\RA schemes based on consecutive self-measurements~\cite{erasmus,ibrahim2017seed} attempt to detect transient malware 
that comes and goes between two successive \RA measurements. The strategy is for \prv to intermittently (based on an either 
periodic or unpredictable schedule) and unilaterally invoke its \RA functionality. Then, either \prv self-reports to 
\vrf~\cite{ibrahim2017seed}, or it accumulates measurements locally and waits for \vrf to explicitly request them \cite{erasmus}.
Upon receiving \RA response(s), \vrf checks for malware presence at the time of each \RA measurement.
Time intervals used in these \RA schemes are depicted in Figure~\ref{fig:periodic}.
\begin{figure}[ht]
\center
\resizebox{0.8\columnwidth}{!}{ 
	\usetikzlibrary{patterns}
\begin{tikzpicture}[node distance=1.5cm, >=stealth]
\tikzset{
  shbox/.style={
    draw,
    rectangle,
    fill={rgb:black,1;white,2}, 
    pattern=north west lines,
    minimum height=.5cm,
    minimum width=.5cm,
    align=center
  },
  obox/.style={
    draw,
    rectangle,
    fill={rgb:black,1;white,2}, 
    minimum height=.5cm,
    minimum width=.5cm,
    align=center
  }
}

	\coordinate (origin) at (0,0);
	\coordinate (xmax) at (9,0);
	\coordinate (ymax) at (0,3);
	\draw[thick, ->] (origin) -- (xmax) ;  
      
    \draw (9,0) node[below=3pt] {Time};
    
	\draw[thick] (origin) -- (ymax);
	
    \draw (0,1) node[left=3pt] {\small \shortstack{Memory\\ Integrity}};
    \draw (0,2) node[left=3pt] {\small \shortstack{Attest. \\ Compute.}};
    
    \draw [fill={rgb:black,1;white,2}, pattern=north west lines] (1,1.7) rectangle (2.5,2.3);
    \draw [fill={rgb:black,1;white,2}, pattern=north west lines] (3.5,1.7) rectangle (5,2.3);
    \draw [fill={rgb:black,1;white,2}, pattern=north west lines] (6,1.7) rectangle (7.5,2.3);
    
    \draw [fill={rgb:black,1;white,2}] (1,.7) rectangle (2.5,1.3);
    \draw [fill={rgb:black,1;white,2}] (3.5,.7) rectangle (5,1.3);
    \draw [fill={rgb:black,1;white,2}] (6,.7) rectangle (7.5,1.3);
    
    \draw [dashed] (1,0) -- (1,3);
    \draw [dashed] (2.5,0) -- (2.5,3);
    \draw [dashed] (3.5,0) -- (3.5,3);
    \draw [dashed] (5,0) -- (5,3);
    \draw [dashed] (6,0) -- (6,3);
    \draw [dashed] (7.5,0) -- (7.5,3);
    
    \draw [line width=0.2mm] (0.5,3.25) -- (8,3.25);
    \draw [line width=0.2mm, ->] (0.5,3.25) -- (0.5,2.75);
    \draw [line width=0.2mm, ->] (3,3.25) -- (3,2.75);
    \draw [line width=0.2mm, ->] (5.5,3.25) -- (5.5,2.75);
    \draw [line width=0.2mm, ->] (8,3.25) -- (8,2.75);
    \draw [line width=0.2mm] (4.25,3.25) -- (4.25,3.5);
    \draw (4.25,3.75) node {Vulnerability Windows};
    
    \draw (.5,-0.25) node {$C_{app}$};
    \draw (3,-0.25) node {$C_{app}$};
    \draw (5.5,-0.25) node {$C_{app}$};
    \draw (8,-0.25) node {$C_{app}$};
    
    \draw (1.75,-0.25) node {$C_{\RA}$};
    \draw (4.25,-0.25) node {$C_{\RA}$};
    \draw (6.75,-0.25) node {$C_{\RA}$};
    
    \foreach \x in {1,2.5,3.5,5,6,7.5}
    	\draw (\x cm, 3pt) -- (\x cm, -3pt);

\end{tikzpicture}
}
	\caption{Consecutive Self-Measurements}
	\label{fig:periodic}
	\vspace{-2mm}
\end{figure}
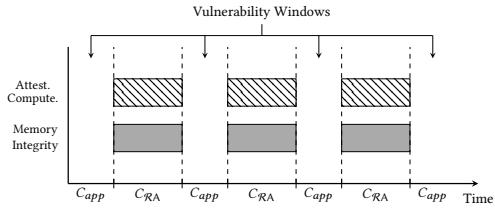

Note that consecutive measurements always leave time gaps during which transient malware presence would not be detected. 
The only way to detect all transient malware with self-measurement schemes is to invoke \RA functionality on \prv with a
sufficiently high frequency, such that the fastest possible transient malware can not come and go undetected. 
However, even if it were easy (which it is not) to determine such ``sufficiently high frequency'',
doing so is horrendously costly, as we show below. We define CPU utilization ($U$) in 
a consecutive scheme as the percentage of CPU cycles that can be used by a regular application ($C_{app}$), i.e, 
cycles other than those spent on self-measurements ($C_{\RA}$):
\begin{equation}
U = \frac{C_{app}}{C_{app}+C_{\RA}}
\end{equation}
As discussed above, guaranteed detection of transient malware via consecutive self-measurements requires that:
\begin{equation}\label{eq:applessadv}
C_{app} < C_{\sadv}
\end{equation}
where $C_{\sadv}$ is the hypothetical number of instruction cycles used by the fastest transient malware, capable of 
infecting \prv, performing its tasks, and erasing itself. To illustrate this point, we 
assume a conservative number for $C_{\sadv}$ to be $10^6$ cycles. In this case:
\begin{equation}
C_{\sadv} = 10^6 \implies
C_{app} < 10^6 \implies
U < \frac{10^6}{10^6+C_{\RA}}\\
\end{equation}
For example with $C_{\RA}$, consider the number of CPU cycles required by \vrased 
(other hybrid \RA architectures, e.g.,~\cite{smart}, have similar costs) to attest a program memory of 
$4$KB: $C_{\RA} = 3.6 \times 10^6$ CPU cycles (about half a second in a typical $8$MHz low-end MCU).
\begin{equation}
U < \frac{10^6}{10^6+3.6 \times 10^6} \implies U < 21.74\%\\
\end{equation}
To detect transient malware, a large fraction of CPU cycles (almost 80\% in this toy example) is spent on \RA computation. 
In practice, it is hard to determine $C_{\sadv}$ and, in some cases (e.g., changing a general-purpose 
input/output value to trigger actuation), it is likely to be much lower than $10^6$ cycles, resulting in even lower CPU utilization 
left for legitimate applications running on \prv. Therefore, detection of all transient malware using consecutive 
self-measurements is impractical. This also applies to the case where the interval between successive 
measurements is variable and/or randomly selected from a range [$0$, $t_{max}$]. As discussed 
in~\cite{ibrahim2017seed}, this is because  it must be that $t_{max}<C_{\adv}$ in order to achieve 
negligible probability of malware evasion.

\begin{figure}[ht]
\center
\resizebox{0.8\columnwidth}{!}{ 
	\usetikzlibrary{patterns}
\begin{tikzpicture}[node distance=1.5cm, >=stealth]
\tikzset{
  shbox/.style={
    draw,
    rectangle,
    fill={rgb:black,1;white,2}, 
    pattern=north west lines,
    minimum height=.5cm,
    minimum width=.5cm,
    align=center
  },
  obox/.style={
    draw,
    rectangle,
    fill={rgb:black,1;white,2}, 
    minimum height=.5cm,
    minimum width=.5cm,
    align=center
  }
}

	\coordinate (origin) at (0,0);
	\coordinate (xmax) at (9,0);
	\coordinate (ymax) at (0,3);
	\draw[thick, ->] (origin) -- (xmax) ;  
      
    \draw (9,0) node[below=3pt] {Time};
    
	\draw[thick] (origin) -- (ymax);
	
    \draw (0,1) node[left=3pt] {\small \shortstack{Memory\\ Integrity}};
    \draw (0,2) node[left=3pt] {\small \shortstack{Attest. \\ Compute.}};
    
    \draw [fill={rgb:black,1;white,2}, pattern=north west lines] (1,1.7) rectangle (2.5,2.3);
    \draw [fill={rgb:black,1;white,2}, pattern=north west lines] (6,1.7) rectangle (7.5,2.3);
    
    \draw [fill={rgb:black,1;white,2}] (0,.7) rectangle (7.5,1.3);
    
    \draw [dashed] (1,0) -- (1,3);
    \draw [dashed] (2.5,0) -- (2.5,3);
    \draw [dashed] (6,0) -- (6,3);
    \draw [dashed] (7.5,0) -- (7.5,3);
    
    \draw [line width=0.2mm] (0.5,3.25) -- (8,3.25);
    \draw [line width=0.2mm, ->] (0.5,3.25) -- (0.5,2.75);
    \draw [line width=0.2mm, ->] (4.25,3.25) -- (4.25,2.75);
    \draw [line width=0.2mm, ->] (8,3.25) -- (8,2.75);
    \draw [line width=0.2mm] (4.25,3.25) -- (4.25,3.5);
    \draw (4.25,3.75) node {\toctou-Security};
    
    \draw (.5,-0.25) node {$C_{app}$};
    \draw (4.25,-0.25) node {$C_{app}$};
    \draw (8,-0.25) node {$C_{app}$};
    
    \draw (1.75,-0.25) node {$C_{\RA}$};
    \draw (6.75,-0.25) node {$C_{\RA}$};
    
    \foreach \x in {1,2.5,6,7.5}
    	\draw (\x cm, 3pt) -- (\x cm, -3pt);

\end{tikzpicture}
}
	\caption{\toctou-Secure \RA}
	\label{fig:toctouless}
	\vspace{-2mm}
\end{figure}

As shown in Figure~\ref{fig:toctouless}, \toctou-Secure \RA (per Definition~\ref{def:toctou_sec}) allows \vrf to 
ascertain memory integrity independently from the time between successive \RA measurements, regardless of 
transient malware's speed. In the next sections, we propose two \toctou-Secure techniques
and show their security with respect to Definition~\ref{def:toctou_sec}.

\section{\acroa: RTC-Based TOCTOU-Secure Technique}\label{sec:rtc_rata}
In hybrid \RA, trusted software (\sw) is usually responsible for generating the authenticated \RA response (\utoken) 
and all semantic information therein. Meanwhile, trusted hardware (\hw) is responsible for ensuring that \sw executes 
as expected, preventing leakage of its cryptographic secrets, and handling unexpected or malicious behavior 
during execution. To address \toctou, we propose a paradigm shift by allowing (formally verified) \hw to also 
provide some context about \prv's memory state. 

We now overview \acroa -- a simple technique 
that requires \prv to have a reliable read-only Real-Time Clock (RTC) synchronized with \vrf.
However, RTCs are not readily available on low-end MCUs and secure clock synchronization in distributed systems 
is challenging~\cite{anwar2019applications,annessi2017s,narula2018requirements}, especially for low-end embedded 
systems~\cite{du2008security,ganeriwal2005secure}. Nonetheless, 
we start with this simple approach to show the main idea behind \toctou-Secure \RA.
Next, Section~\ref{sec:clockless} proposes an alternative variant that removes the RTC requirement, as long as \vrf 
requests are authenticated by \prv. Note that \vrf authentication is already included in some current 
hybrid \RA architectures, including \vrased.

\subsection{\acroa: Design \& Security}
\begin{figure}[!ht]
\centering
\resizebox{0.7\columnwidth}{!}{%
	\begin{tikzpicture}[node distance=1.5cm, >=stealth]
\tikzset{
  msp430/.style={
    draw,
    rectangle,
    minimum height=5cm,
    minimum width=8cm,
    align=center
  },
  membb/.style={
    draw,
    rectangle,
    minimum height=9cm,
    minimum width=1.5cm,
    align=center
  },
  mem/.style={
    draw,
    rectangle,
    minimum height=.5cm,
    minimum width=1.5cm,
    align=center
  },
  smallbox/.style={
    draw,
    rectangle,
    minimum height=1.8cm,
    minimum width=2.8cm,
    align=center
  },
  tinybox/.style={
    draw,
    rectangle,
    minimum height=.8cm,
    minimum width=1.8cm,
    align=center
  },
  hwmod/.style={
    draw,
    rectangle,
    minimum height=5cm,
    minimum width=4cm,
    align=center
  },
  chalbox/.style={
    draw,
    fill={rgb:black,1;white,2},
    rectangle,
    minimum height=.8cm,
    minimum width=2cm,
    align=center
  },
  lmtbox/.style={
    draw,
    fill={rgb:black,1;white,2},
    rectangle,
    minimum height=.8cm,
    minimum width=3cm,
    align=center
  },
  erbox/.style={
    draw,
    rectangle,
    minimum height=4cm,
    minimum width=2cm,
    align=center
  },
  orbox/.style={
    draw,
    rectangle,
    minimum height=3cm,
    minimum width=2cm,
    align=center
  }
}
	\node[msp430] (msp) at (0,0) {\huge MCU CORE};
        \node[smallbox, below right = 4cm of msp.south] (vrased) {\Large \vrased};
        \node[tinybox, below = .7cm of vrased.south] (vape) {\Large RATA};
        \node[hwmod] at (4,-6.5) (hwmod) {};
        \node[overlay,anchor=north east] at (hwmod.north east) (test) {\texttt{\huge HW-Mod}};
        
        \path (msp.south) +(-1,0) coordinate (msp_0);
        \path (msp.south) +(-2.5,0) coordinate (msp_1);
        \path (vrased.west) +(0,.5) coordinate (vrased_0);
        \path (vrased.west) +(0,-.5) coordinate (vrased_1);
        \draw[thick, ->] (msp_1) |- (vrased_1); 
        \node[inner sep=5pt,right, fill=white,draw] at ($(msp_1) + (-.75,-2)$) {\shortstack{$PC$,\\$irq$,\\$R_{en}$,\\$W_{en}$,\\$D_{addr}$,\\$\dmaen$,\\$\dmaaddr$}};
        \draw[thick, <-] (msp_0) |- (vrased_0) ; 
        \node[inner sep=5pt,right, fill=white, draw] at ($(vrased_0) + (-3,0)$) {$reset$};
        
        \node[lmtbox, below left = 1.5cm and 8cm of hwmod.west] (lmt) {$LMT$};
        
        \path (vape.west) +(0,0) coordinate (vape_top);
        \path (vape.west) +(0,-.3) coordinate (vape_bot);
        \path (lmt.east) +(0,-.15) coordinate (lmt_bot);
        \path (lmt.east) +(0,.15) coordinate (lmt_top);
        \draw[thick, ->] (vape_top) -- (lmt_top); 
        \draw[thick, ->] (lmt_bot) -- (vape_bot); 
        
		\path (vrased_1) +(-.5, 0) coordinate (vrased_dot);
        \path (vape.west) +(0,0.2) coordinate (vape_connect);
        \draw[thick, ->] (vrased_dot) |- (vape_connect); 
        
        \foreach \n in { vrased_dot}
			\node at (\n)[circle,fill,inner sep=1.5pt]{};
		
        \path (lmt.north west) +(0,5.5) coordinate (memleft_start);
        \path (lmt.south west) +(0,0) coordinate (memleft_end);
        \draw[thick] (memleft_start) -- (memleft_end); 
        \path (lmt.north east) +(0,5.5) coordinate (memright_start);
        \path (lmt.south east) +(0,0) coordinate (memright_end);
        \draw[thick] (memright_start) -- (memright_end); 
        \draw[thick] (memright_start) -- (memleft_start); 
        \draw[thick] (memright_end) -- (memleft_end); 
        
		\path (lmt.south) +(0, 4.2) coordinate (mem_name);
        \node[overlay,anchor=north] at (mem_name) (test) {\shortstack{\Large Program \\ \Large Memory}};
        
\end{tikzpicture}
}
\caption{\acro module in the overall system architecture}\label{fig:sys-arch}
\vspace{-3mm}
\end{figure}
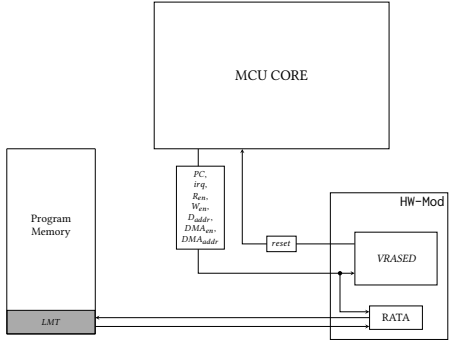

\acroa is illustrated in Figure~\ref{fig:sys-arch}; it is designed as a verified hardware module behaving as follows:

 \noindent~\quad {\bf (1)}  It monitors a set of CPU signals and detects whenever any location within $AR$ is written. 
 This is achieved by checking the value of signals $D_{addr}$, $W_{en}$, $DMA_{addr}$, and $DMA_{en}$ 
 (see Section~\ref{sec:background}). These signals allow for detection of memory modifications either by 
 CPU or by DMA.
 
 \noindent~\quad {\bf (2)} Whenever a modification to $AR$ is detected, \acroa logs the timestamp by reading the current time 
 from  the RTC and storing it in a fixed memory location, called Latest Modification Time ($LMT$).
 
 \noindent~\quad {\bf (3)} In the memory layout, $LMT \in AR$. Also, \acroa enforces that LMT is always read-only for 
 all software executing on the MCU, and for DMA.\\

\begin{figure*}
\begin{mdframed}
\vspace{-1em}
\footnotesize
\begin{construction}[\acroa]\label{def:cons_RTC}
Suppose $LMT$ is a program memory region within $AR$ ($LMT \in AR$):\\
	\begin{itemize}
		\item $\request^{\vrf \rightarrow \dev}\OParan \CParan$: \vrf generates a random $l$-bits challenge $\chal \leftarrow \$\{0,1\}^l$ 
		and sends it to \dev.
		\item $\attest^{\dev \rightarrow \vrf}\OParan \chal \CParan$:
		Upon receiving \chal, \prv calls \vrased \sw's \RA function to compute $\utoken = HMAC(KDF(\attkey, \chal), AR)$ 
		and sends $t_{LMT}||\utoken$ to \vrf, where $t_{LMT}$ is the value stored in $LMT$.\\
		At all times, \acroa hardware in \prv enforces the following invariants:
		
		-- $LMT$ is read-only to software:
		\begin{equation}\label{eq:prop1}
        		\textbf{Formal statement (LTL): }\quad{\bf G}\{ Mod\_Mem(LMT) \rightarrow reset \}
		\end{equation}
		-- $LMT$ is overwritten with the current time from RTC if, and only if, $AR$ is modified:
		\begin{equation}\label{eq:prop2}
        		\textbf{Formal statement (LTL): } \quad {\bf G}\{Mod\_Mem(AR) \leftrightarrow set_{LMT}\}
		\end{equation}
		where $reset$ is a 1-bit signal that triggers an immediate reset of the MCU, and $set_{LMT}$ is a 1-bit output 
		signal of \acroa controlling the value of $LMT$ reserved memory. Whenever $set_{LMT} = 1$, $LMT$ is 
		updated with the current value from the real-time clock (RTC). LMT maintains its previous value otherwise.
		\item $\vrfy^{\vrf}\OParan \utoken, \chal, M, t_0, t_{LMT} \CParan$:
		$t_0$ is an arbitrary time chosen by \vrf, as in Definition~\ref{def:toctou_sec}. Upon receiving $t_{LMT}||\utoken$ \vrf checks:
		 \begin{equation}
		  t_{LMT} < t_0
		 \end{equation}
         \begin{equation}\label{eq:condition_att}
            \utoken \equiv \hmac(KDF(\attkey, MR), M)
         \end{equation}
        where $M$ is the expected value of $AR$ reflecting $LMT=t_{LMT}$, 
        as received from \prv.
        \vrfy returns $1$ if and only if both checks succeed.
	\end{itemize}
\end{construction}
\end{mdframed}
\end{figure*}

Note that, by enforcing $LMT \in AR$, the  attestation result 
$\utoken=HMAC(KDF(\attkey, MR),AR)$ includes the authenticated value of $LMT$ --  the time corresponding 
to the latest modification of $AR$. As part of the \vrfy algorithm, \vrf compares this information with the 
time of the last authorized modification ($t_0$ of Definition~\ref{def:toctou_sec}) of  $AR$ to check whether any 
unauthorized modifications occurred since then. The general idea is further specified in Construction~\ref{def:cons_RTC}, 
which shows how \acroa can be seamlessly integrated into \vrased, enforcing two additional properties in hardware to 
obtain \toctou-Security. These properties are formalized in LTL in Equations~\ref{eq:prop1} and~\ref{eq:prop2} 
of Construction~\ref{def:cons_RTC}.

\begin{figure}[t]
	\centering
	\includegraphics[width=1.1\columnwidth]{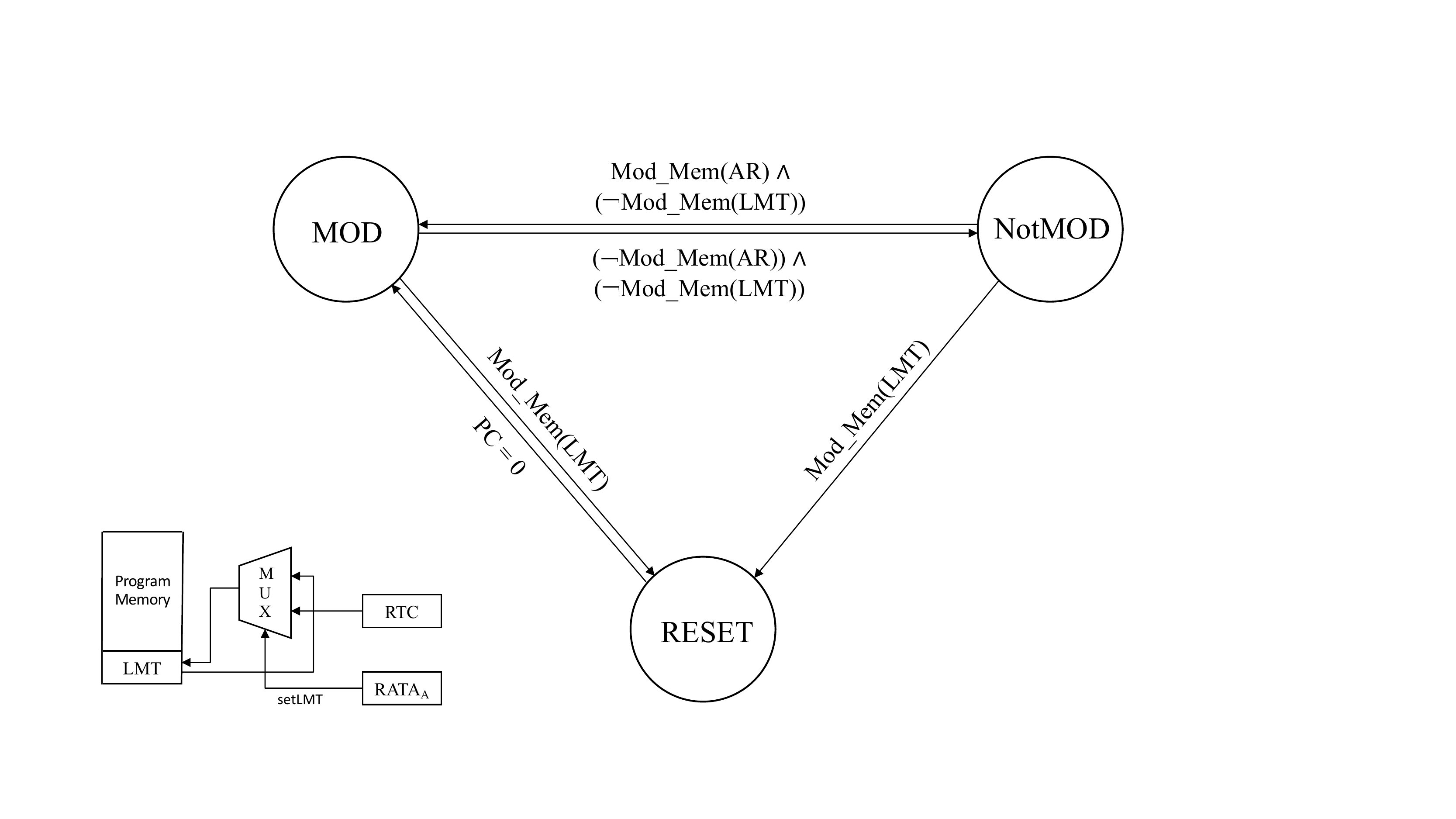}
	\caption{\acroa FSM for RTC-based \toctou-secure \RA}\label{fig:fsm}
	\vspace{-3mm}
\end{figure}

We show that Construction~\ref{def:cons_RTC} is secure as long as \acroa implementation adheres to LTL statements 
in Equations~\ref{eq:prop1} and~\ref{eq:prop2}. This verification is discussed in Section~\ref{sec:imp}. 
The cryptographic proof is by reduction from \vrased security (per Definition~\ref{def:vrased_sec}) to \toctou-Security 
(per Definition~\ref{def:toctou_sec}) of Construction~\ref{def:cons_RTC}.  For its part, \vrased is shown secure 
according to Definition~\ref{def:vrased_sec} as long as \hmac is a secure, i.e., existentially unforgeable~\cite{unforge}, 
MAC (see~\cite{vrasedp} for details). The proof of Theorem~\ref{th:toctou_sec} is presented in Appendix~\ref{apdx:proof_rataa}.

\begin{mdframed}
\vspace{-1em}
	\textit{\begin{theorem}\label{th:toctou_sec}
			\small
			Construction~\ref{def:cons_RTC} is \toctou-Secure according to Definition~\ref{def:toctou_sec} as 
			long as \vrased is secure according to Definition~\ref{def:vrased_sec}.
	\end{theorem}}
\end{mdframed}

\subsection{\acroa: Implementation \& Verification}\label{sec:imp}
Construction~\ref{def:cons_RTC} (and respective security proof) assumes that properties in 
Equations~\ref{eq:prop1} and~\ref{eq:prop2} are enforced by \acroa.
Figure~\ref{fig:fsm} shows a formally verified FSM corresponding to this implementation.
It enforces two properties of Equations~\ref{eq:prop1} and~\ref{eq:prop2}.
This FSM is implemented as a Mealy machine, where output changes anytime based on both the current state and current input values.
The FSM takes as an input a subset of signals, shown in Figure~\ref{fig:sys-arch},
and produces two 1-bit outputs: $reset$ to trigger an immediate reset and
$set_{LMT}$ to control the value of $LMT$ memory (see Construction~\ref{def:cons_RTC}).
$reset$ is 1 whenever FSM transitions to $RESET$ state and while it remains in that state; it remains $0$ otherwise.
Whereas, $set_{LMT}$ is $1$ when FSM transitions to $MOD$ state, and becomes $0$ whenever it transitions 
out of $MOD$ state. $set_{LMT}=0$ in all other cases.

The FSM works by monitoring write access to $LMT$ and 
transitioning to $RESET$ whenever such attempt happens.
When the system is running (i.e., $reset=0$), FSM also monitors write access to $AR$ and 
transitions to $MOD$ state whenever it happens. The FSM transitions back to $NotMOD$ state if $AR$ is not being modified.
We design the FSM in Verilog HDL and automatically translate into SMV using Verilog2SMV~\cite{irfan2016verilog2smv}. 
Finally, we use NuSMV model checker~\cite{nusmv} to prove that the FSM complies with
invariants~\ref{eq:prop1} and~\ref{eq:prop2}. The implementation and correspondent verification are available in~\cite{repo}.

\noindent\emph{\textbf{Remark 3:} Since deletion is a ``write'' operation, malware can not erase itself at 
runtime without being detected by \acro. Conversely, any attempt to reprogram flash ($AR$) directly via wired 
connection requires device re-initialization. Both \acroa/\acrob always update $LMT$ on initialization/reset/reboot. 
Hence, these modifications are also detected.}

\noindent\emph{\textbf{Remark 4:} The ability to cause a reset by attempting to write to LMT yields no 
advantage for \adv, since any bare-metal software (including malware) can always trigger a reset on an 
unmodified low-end device, e.g., by inducing software faults.}

\section{\acrob: Clockless TOCTOU-Secure \RA Technique}\label{sec:clockless}
We now describe \acrob: a \toctou-Secure technique that requires no clock on \prv.
We apply the ideas from \acroa by using hardware to convey authenticated information about \emph{the time of the 
latest memory modification} as part of the attestation result. However, lack of RTC prevents any notion of ``time'' on \prv's end. 
To cope with this, we rely on \vrf to convey information tied to a given point in time, according to \vrf's own local clock. This is done 
as a part of \RA \request algorithm. In fact, \acrob uses the attestation challenge (\chal) itself in this task, taking advantage of the 
fact that \chal is \textit{unique} per \request and is available in any \RA technique, thus incurring no additional communication overhead. 
Security of \acrob is tightly coupled with authentication of \vrf \request, which is already part of \vrased architecture~\cite{vrasedp}; 
see Appendix~\ref{apdx:vrased_auth} for details.

\subsection{\acrob~-- Design \& Security}
\begin{figure*}
	\begin{mdframed}
	\vspace{-1em}
	\footnotesize
		\begin{construction}[\acrob]\label{def:cons_clockless}
			Suppose $LMT$ is a memory region within $AR$ (i.e., $LMT \in AR$) and $P$ is a challenge-time association pair, stored by \vrf. Initially $P=(\perp,\perp)$.
			\acrob is specified as follows:
            \vspace{1mm}
			\begin{compactitem}
				\item $\request^{\vrf \rightarrow \dev}\OParan \CParan$: \vrf generates a pair $[\chal,\auth]$ 
				according to \vrased authentication algorithm (see Appendix~\ref{apdx:vrased_auth} for details) and sends it \prv.
				\vspace{1mm}
				\item $\attest^{\dev \rightarrow \vrf}\OParan \chal , \auth \CParan$: 
					Upon receiving $[\chal,\auth]$, \prv behaves as follows: 
					\begin{enumerate}
						\item Call\ \vrased \sw's \RA function to use \auth to authenticate \chal. 
							If authentication succeeds, proceed to next step. Otherwise, ignore the request.
						\item Compute $\utoken = HMAC(KDF(\attkey, \chal), AR)$, restricted that $LMT \in AR$, where $|LMT| = |\chal|$.
						\item Send $LMT||\utoken$ to \vrf.
					\end{enumerate}
				To support this operation, at all times, \acrob hardware on \prv enforces the following:
				
				-- LMT is read-only to software:
				\begin{equation}\label{eq:2prop1}
				\textbf{Formal statement (LTL): }\quad{\bf G}\{Mod\_Mem(LMT) \rightarrow reset\}
				\end{equation}

				-- LMT is never updated without authentication:
				\begin{equation}\label{eq:2prop2}
				\textbf{Formal statement (LTL): } \quad{\bf G}\{[\neg UP_{LMT} \land {\bf X}(UP_{LMT})] \rightarrow 
				{\bf X}(PC = CR_{auth})\}
				\end{equation}
				-- Modification(s) to $AR$ imply updating LMT in the next authenticated \attest call:
				\begin{equation}\label{eq:2prop3}
				\textbf{Formal statement (LTL):}\quad{\bf G}\{Mod\_Mem(AR)\lor reset \rightarrow [(PC = CR_{auth} 
				\rightarrow UP_{LMT})~{\bf W}~ (PC = CR_{max} \lor reset)]\}
				\end{equation}
				where $reset$ is a 1-bit signal that triggers an immediate reset of the MCU, and $UP_{LMT}$ is a 
				1-bit signal that, when set to $1$, replaces the content of $LMT$ with the current value stored in 
				$MR$ region (i.e., \chal). $LMT$ maintains its previous value otherwise.
				\vspace{1.5mm}
				\item $\vrfy^{\vrf}\OParan \utoken, \chal, M, t_0, P , LMT\CParan$: 
					Let $t_0$ denote a time chosen by \vrf, as in Definition~\ref{def:toctou_sec}.
					Denote the current values in the challenge-time association pair stored by \vrf as $P=(\chal_P, t_{P})$.
					Upon receiving $LMT||\utoken$, \vrf behaves as follows:
				\begin{enumerate} 
                    \item Check if $\utoken \equiv \hmac(KDF(\attkey, \chal), M)$, where $M$ is the expected $AR$ value.
					Since $AR$ includes $LMT$, $M$ is set to contain the value of $LMT$, as received from \prv. 
					Hence, this checks also assures integrity of $LMT$ in $AR$.
					If this check fails, {\bf return 0}, otherwise, proceed to step 2;
					\item If $LMT=\chal_P$ and $t_0>t_P$, {\bf return 1}, otherwise, proceed to step 3;
					\item Set $P=(LMT,current\_time)$ and {\bf return 0};\\
				\end{enumerate}
			\end{compactitem}
		\end{construction}
	\end{mdframed}
\end{figure*}

The design of \acrob remains consistent with Figure~\ref{fig:sys-arch}.
\acrob monitors the same set of MCU signals as \acroa and also works by overwriting the special 
memory region $LMT \in AR$. However, instead of logging an RTC timestamp to $LMT$, it logs \chal, 
which was sent by \vrf as a part of its \request and given as input to $\attest(\chal, ...)$. $LMT$ is overwritten 
with the currently received \chal if and only if, a modification of $AR$ occurred since the previous \attest instance. 
In summary, \acrob security relies on the following properties, enforced by its verified hardware implementation 
(see Section~\ref{sec:imp_clockless}):

\noindent~\quad{\bf (1)} Similar to \acroa, no software running on \prv can overwrite $LMT$, i.e., $LMT$ 
is only modifiable by \acrob hardware.

\noindent~\quad{\bf (2)} An update to $LMT$ is triggered only immediately after a successful authentication 
during \attest computation.

\noindent~\quad{\bf (3)} The first successful authentication happening after a modification of $AR$ always 
causes $LMT$ to be updated with the current value of $\chal$ which is stored in $MR$. (Recall from 
Table~\ref{tab:notation} that $MR$ is the memory location from which \attest reads the value of \chal.)\\

Let $\chal_1$ and $\utoken_1$ denote the attestation challenge and response successfully sent/received by \vrf, in a given \RA interaction.
\vrf interprets \RA results as follows: if $\utoken_1$ is a valid response, i.e., it corresponds to an expected $AR$ value,
time $t_1$ when such response is received is saved locally by \vrf, associated to $\chal_1$. In subsequent attestation results 
($\utoken_{2}$, $\utoken_{3}$, ...), \vrf checks the value of $LMT$ for correspondence with $\chal_1$. If $LMT \neq \chal_1$, 
\vrf learns that $AR$ was modified after $t_1$.
This stems from \acrob verified module, which guarantees that $LMT$ is 
always overwritten with the newly received challenge if a \toctou happens between consecutive calls to \attest.
In this design, we highlight the following observations:
 
 \noindent~\quad-- \textbf{Authentication of \vrf \request} is instrumental to \acrob security. Without it, \sadv can simply choose 
 $\chal_{\sadv}$ and call $\attest(\chal_{\sadv})$ after an unauthorized modification of $AR$, thus setting $LMT=\chal_{\sadv}$ of 
 its choice. By choosing $\chal_{\sadv}$ as a value previously used by \vrf, \sadv can easily convince \vrf that no \toctou occurred
 between measurements. In other words, lack of \request authentication allows \adv to modify $LMT$ at will, rendering write 
 protection of $LMT$ useless.
 
 \noindent~\quad-- \textbf{Uniqueness of $LMT$} must be enforced, e.g., by having \vrf randomly sample \chal from a sufficiently large space or 
 use \chal as a monotonically increasing counter, depending on specifics of \request algorithm. 
 If \chal is reused after $n$ instances of \request, \sadv can wait for the $n$-th 
 authentic \request to complete, infect \dev, perform its tasks, and leave \dev before the 
 $(n+1)$-st \request occurs (with a reused \chal), resulting in a valid response and
  compromised \toctou-Security. For example, if we use $LMT$ as a dirty-bit (instead of \chal), 
  security can be subverted in two \request-s,  even if they are properly authenticated. \\

\acrob is specified in Construction~\ref{def:cons_clockless}.
\prv's hardware module controls the value of a $1$-bit signal $UP_{LMT}$. When set to $1$, $UP_{LMT}$ updates $LMT$ 
with the current value of $MR$; otherwise, $LMT$ maintains its current value.
\acrob hardware detects successful authentication of \vrf by checking whether the program counter $PC$ points to the 
instruction reached immediately after successful authentication. Note that the instruction at location $CR_{auth}$ 
is never reached {\bf unless} authentication succeeds.
Note that, unlike \acroa, \vrf in \acrob learns whether a modification 
occurred since a previous successful attestation response, though not the exact time of that modification.
\acrob security is stated in Theorem~\ref{th:toctou_sec_b}.

\vspace{1mm}
\begin{mdframed}
\vspace{-1em}
\textit{\begin{theorem}\label{th:toctou_sec_b}
\small
Construction~\ref{def:cons_clockless} is \toctou-Secure according to Definition~\ref{def:toctou_sec} as 
long as \vrased is secure according to Definition~\ref{def:vrased_sec}.
\end{theorem}}
\end{mdframed}

Proof of Theorem~\ref{th:toctou_sec_b} is deferred to Appendix~\ref{apdx:proof_ratab}.

\subsection{\acrob: Implementation \& Verification}\label{sec:imp_clockless}
\begin{figure}[t]
	\centering
	\includegraphics[width=1.1\columnwidth]{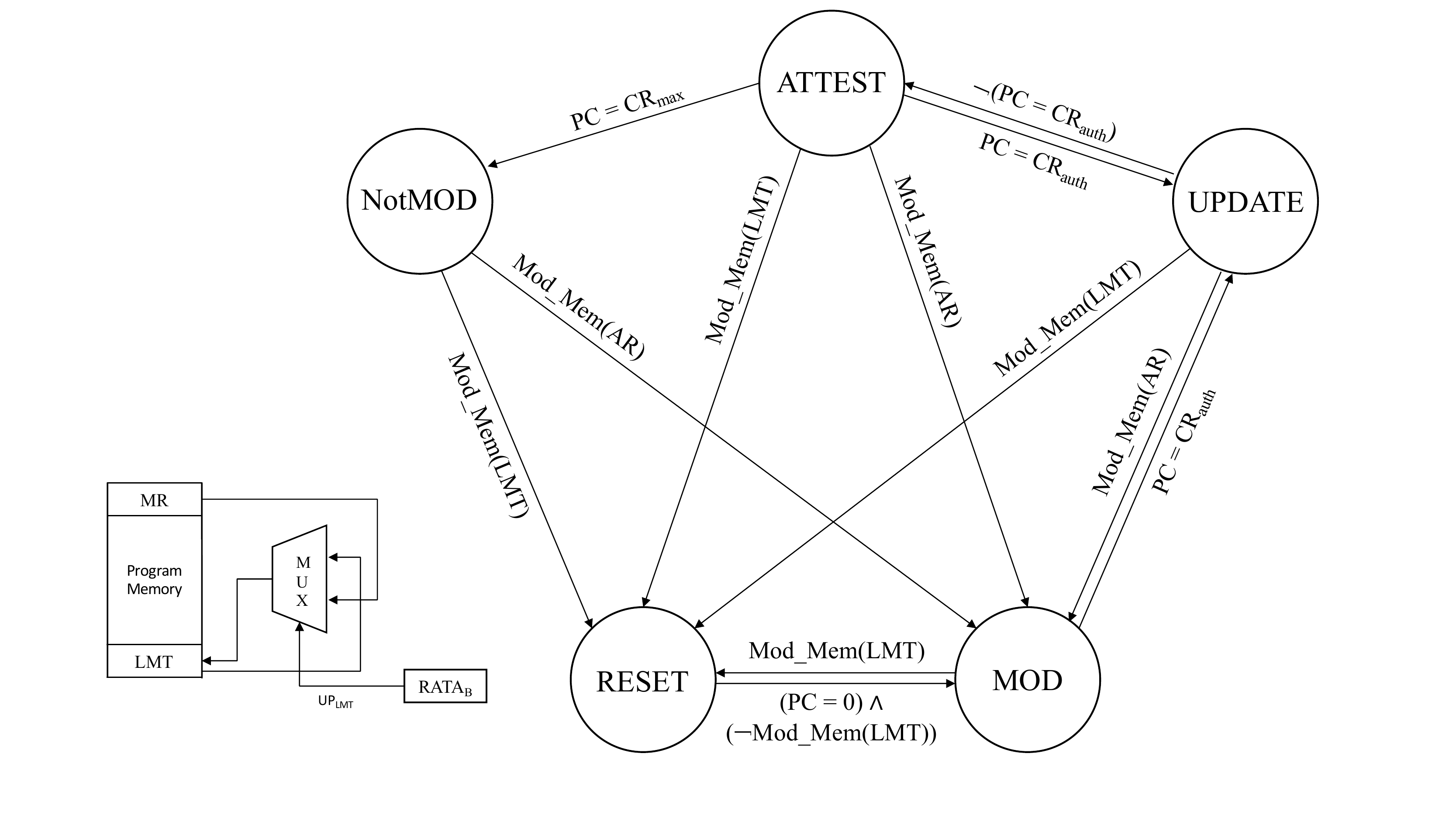}
	\caption{\acrob FSM for clock-less \toctou-secure \RA}\label{fig:clockless_fsm}
	\vspace{-3mm}
\end{figure}
Proof of Theorem~\ref{th:toctou_sec_b} assumes that \acrob hardware adheres to properties in Equations~\ref{eq:2prop1} 
to~\ref{eq:2prop3}. Figure~\ref{fig:clockless_fsm} shows \acrob implementation as an FSM formally verified to adhere to 
these properties. It takes as input a subset of signals, shown in Figure~\ref{fig:sys-arch}  and outputs two 1-bit signals: 
$reset$ triggers an immediate system-wide reset  and $UP_{LMT}$ controls updates to $LMT$ region. $UP_{LMT}=1$ 
whenever the FSM transitions to state $UPDATE$ and has value $0$ in all other states. $reset=1$ whenever the 
FSM transitions to state $RESET$ and remains unchanged while in this state; it remains 0 otherwise.
The FSM operates as follows:
\begin{compactenum}
 \item If a software modification of $LMT$ is attempted, FSM triggers $reset$ immediately, regardless of what state it is in.
 \item If no modifications are made to $AR$ since the previous computation of \attest, FSM remains in $NotMOD$ state.
 \item At any point in time, if a modification to $AR$ is detected, FSM transitions to state $MOD$. This transition indicates that 
 a modification occurred, although it neither alters any output, nor modifies $LMT$. This is because the information to be written 
 to $LMT$ (the value of \chal in the next \request) is not available at this time. 
 \item When a call to \attest is made, two possible actions can occur:
    \begin{enumerate}
    \item If FSM is in $NotMOD$ state, \attest is computed normally and FSM remains in the same state.
    \item Otherwise, FSM stays in $MOD$ state until condition $PC=CR_{auth}$ is met, implying successful authentication of \vrf \request. 
    Then, FSM transitions to state $UPDATE$ causing $UP_{LMT}$ to be set during the transition. Hence, $LMT$ is overwritten 
    with \chal passed as a parameter to the current \attest call. Note that update to $LMT$ happens before the computation of the 
    integrity-ensuring function (\hmac) over $AR$, which happens in state $ATTEST$. 
    Therefore, attestation result \utoken will reflect $LMT=\chal$ as part of $AR$. 
    Once \attest is completed ($PC=CR_{max}$), FSM transitions back to $NotMOD$.
    \end{enumerate}
\end{compactenum}
The same verification tool-chain discussed in Section~\ref{sec:imp} is used to prove that this FSM adheres to 
LTL statements in Equations~\ref{eq:2prop1}, \ref{eq:2prop2}, and~\ref{eq:2prop3}.

\section{Evaluation}\label{sec:eval}
Our prototype is built upon a representative of the low-end class of devices -- TI MSP430 MCU family~\cite{TI-MSP430}.
It extends \vrased (itself built atop OpenMSP430~\cite{openmsp430} -- an open-source implementation of MSP430) 
to enable \toctou detection. It is synthesized and executed using Basys3 commodity FPGA prototyping board.

\noindent\textbf{Hardware Overhead.}
Table~\ref{tab:exp} reflects the analysis of \acro verified hardware overhead.
Similar to some related work~\cite{vrasedp,zeitouni2017atrium,dessouky2017fat,dessouky2018litehax,apex,pure}, we 
consider the hardware overhead in terms of additional LUTs and registers. The increase in the number of LUTs can 
be used as an estimate of the additional chip cost and size required for combinatorial logic, while the number of extra
registers offers an estimate on state registers required by sequential logic in \acro FSMs.
Compared to \vrased, the verified implementation of \acroa module takes $4$ additional registers and $13$ additional LUTs, 
while \acrob increases the number of LUTs and registers by $57$ and $27$, respectively. As far as the unmodified OpenMSP430 
architecture, this represents the overhead of $1.4$\% LUTs and $1.4\%$ registers for \acroa and $3.8\%$ LUTs and 
$4.8\%$ registers for \acrob.

\begin{table}[!htp]
	\resizebox{0.9\linewidth}{!}{ 
		\begin{tabular}{|l|cc|ccc|} \hline\cline{1-6}
			\multirow{2}{*}{Architecture} & \multicolumn{2}{c|}{Hardware} & \multicolumn{3}{c|}{Verification}      \\
                                      & LUT            & Reg           & Verified LoC & Time (s) & Memory (MB) \\ \hline
			OpenMSP430                & 1849           & 692           & -            & -        & -           \\
			\vrased                   & 1862           & 698           & 474          & 0.4      &  13.6       \\
			\acroa                    & 1875           & 702           & 601          & 0.6      &  19.7       \\
			\acrob                    & 1919           & 725           & 656          & 0.8      &  26.1       \\
			\hline\cline{1-6}
		\end{tabular}%
	}
	\centering	\caption{\small Additional hardware and verification cost}
	\label{tab:exp}
\end{table}

\noindent\textbf{Runtime Overhead.} 
\acro does not require any modification to \RA execution. It only ensures that information about the latest 
modification of attested memory is factored into the attestation result.
Hence, it incurs no extra runtime cycles or additional RAM allocation, on top of that of \vrased architecture.
In fact, as we discuss next, in Section~\ref{sec:discussion}, \attest runtime can be reduced to the time 
to attest only LMT. The runtime reduction is presented in Figure~\ref{fig:rata_savings}. This represents a 
reduction of $\approx10$ times compared, e.g., to the number of cycles to attest an $AR$ of size 4KBytes.
The runtime savings increase linearly with the size of $AR$.

\noindent\textbf{Memory Overhead.}
\acroa requires 128-bit of additional storage: 64 bits for RTC and 64 bits for $LMT$.
RTC is implemented using a 64-bit memory cell incremented at 
every clock cycle. This guarantees that RTC does not wrap around during \prv's lifetime since it would take 
more than $70,000$ years for that to happen on MSP430 running at 8MHz and incrementing RTC at every cycle.
In \acroa, $LMT$ is implemented as a 64-bit memory storage and updates its content with RTC value whenever $set_{LMT}$ bit is on.
For \acrob, the memory overhead increases to a total of 512 bits.
256 bits of memory are required by the implementation of \vrased authentication module,
while another 256 bits are used to implement $LMT$ that updates its content with \chal when applicable 
(as described in Section~\ref{sec:clockless}).
This small reserved memory corresponds to 0.1\% of MSP430 memory address space ($64$KBytes in total).

\noindent\textbf{Verification resources.}
We verify \acro on an Ubuntu $18.04$ machine running at $3.40$GHz. Results are shown in Table~\ref{tab:exp}.
\acroa adds $127$ lines of verified Verilog code on top of \vrased. These are needed to enforce $2$ 
invariants in Equations~\ref{eq:prop1} and~\ref{eq:prop2}.
\acrob incurs $182$ additional lines of verified Verilog code, needed to enforce the $3$ invariants in 
Equations~\ref{eq:2prop1},~\ref{eq:2prop2}, and~\ref{eq:2prop1}.
Besides that, \acro verification requires checking existing \vrased invariants.
Overall verification process takes less than one second and consumes at most $26$MB of 
memory, making it suitable for a commodity desktop.

\begin{figure}[t]
	\centering
	\subfigure[Additional LUTs]
	{\includegraphics[width=0.45\columnwidth]{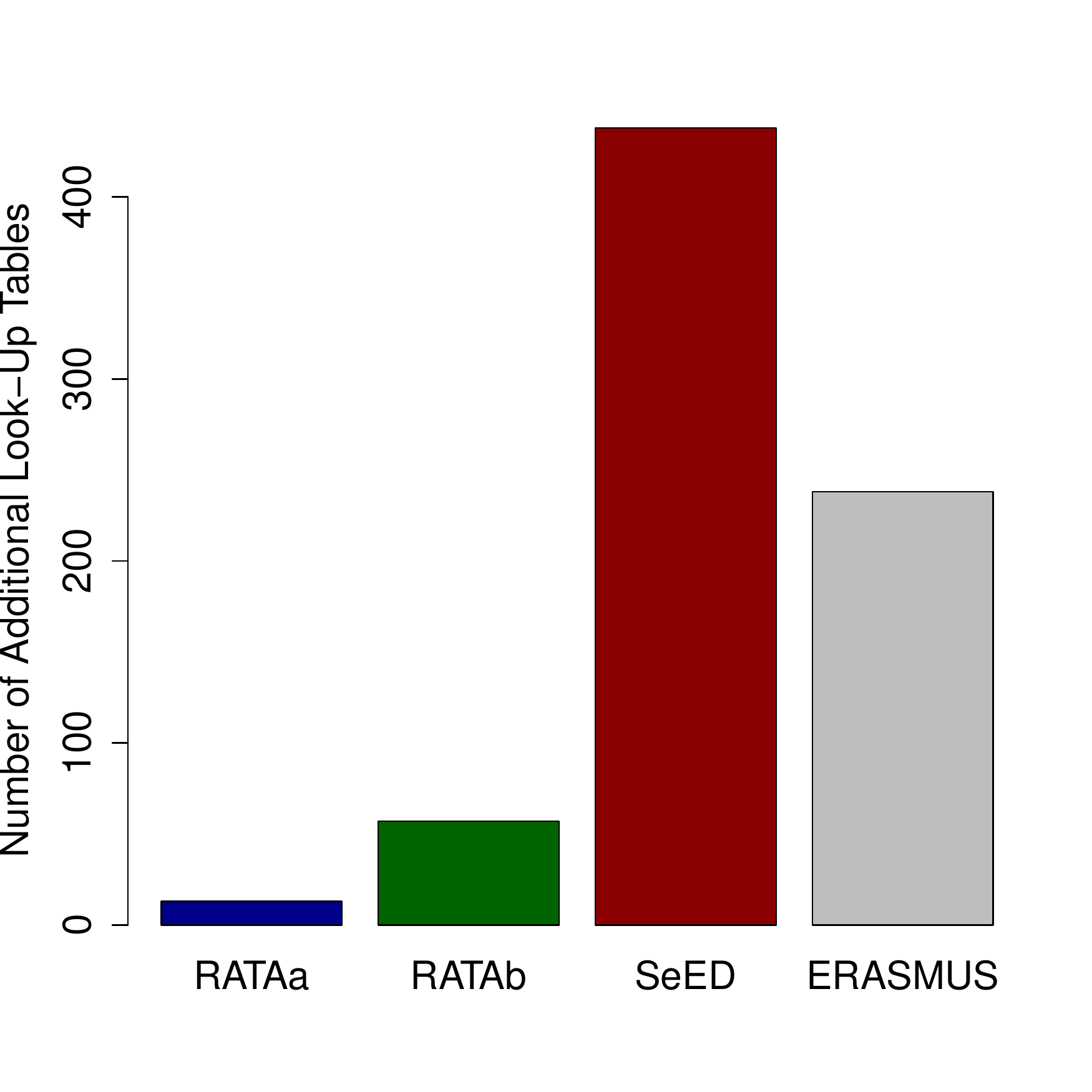}}
	\subfigure[Additional Registers]
	{\includegraphics[width=0.45\columnwidth]{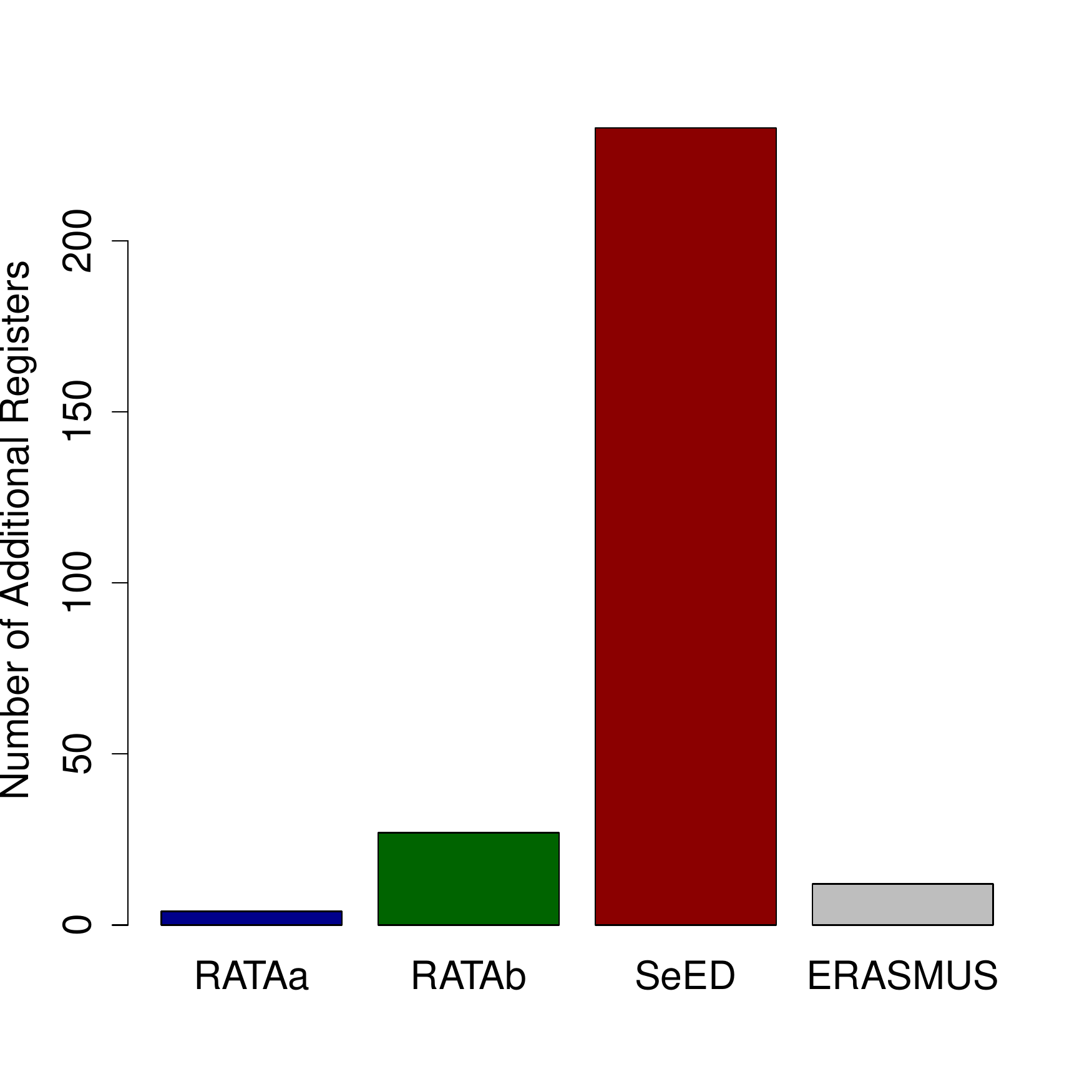}}
	\vspace{-2mm}
	\caption{Hardware overhead. Comparison between \acro and techniques based on self-measurements.}\label{fig:comparison}
	\vspace{-3mm}
\end{figure}
\begin{figure}[t]
	\centering
	\includegraphics[width=0.9\columnwidth]{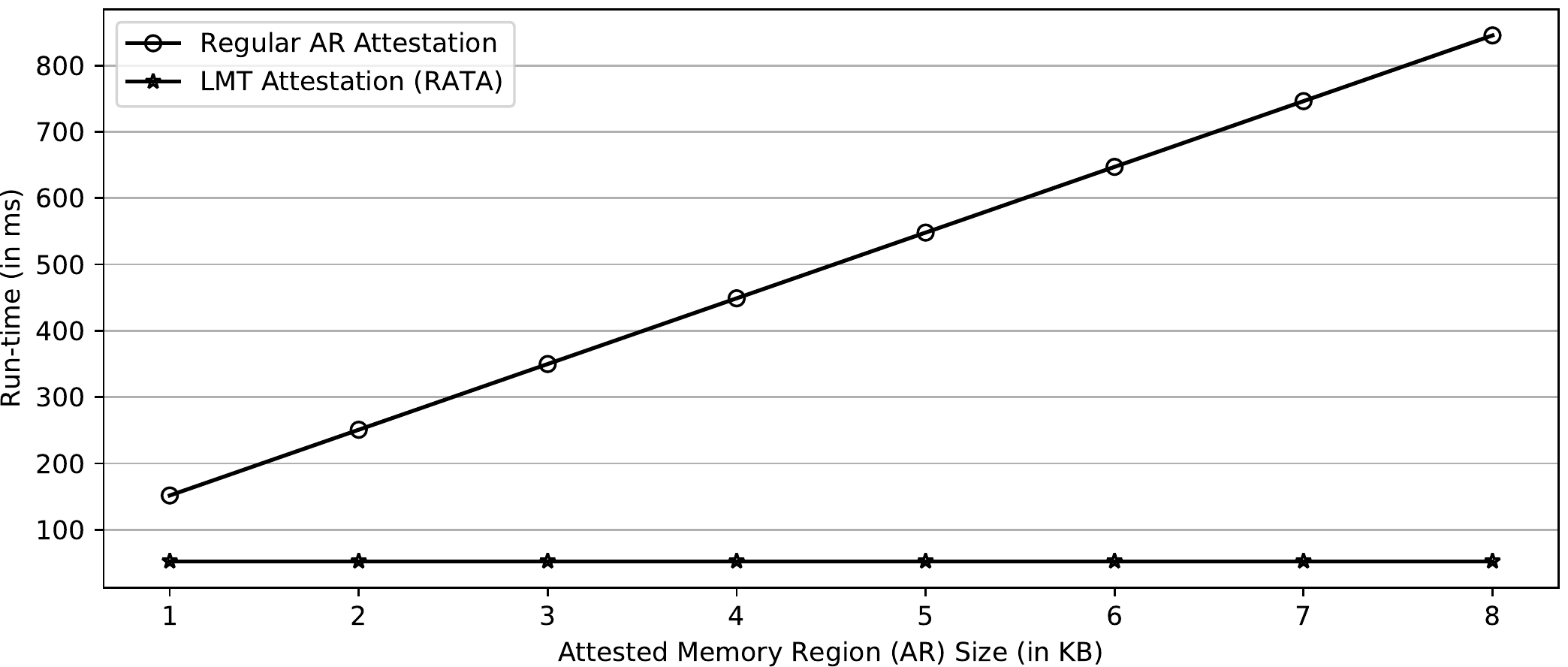}
    \vspace{-2mm}
	\caption{Comparison of $LMT$ attestation time \textbf{Case-1}) with regular attestation of $AR$ (\textbf{Case-2}),
	as a function of $|AR|$. $|LMT|$ is $32$ Bytes. Results on the MSP430 MCU running at 8MHz.}\label{fig:rata_savings}
	\vspace{-3mm}
\end{figure}
\noindent\textbf{Comparison.}
We compare \acro's hardware overhead with that of two recent self-measurement \RA techniques: 
SeED~\cite{ibrahim2017seed} and ERASMUS~\cite{erasmus}.
Even though, as discussed in Section~\ref{sec:motivation}, these techniques do not achieve \toctou-Security 
(per Definition~\ref{def:toctou_sec}), we believe that they are the most closely related approaches to \acro.
SeED extends a $32$-bit Intel architecture, which is higher-end than our target devices, i.e., a 
16-bit TI MSP430. Whereas, ERASMUS was implemented on MSP430.
Figure~\ref{fig:comparison} compares \acro to SeED and ERASMUS  in terms of numbers of additional LUTs and registers.
\acroa require fewer LUTs, compared to both SeED and ERASMUS.
Whereas, \acrob necessitates more registers, compared to ERASMUS, it uses less LUTs than both self-measurements techniques.
In summary, both \acro-s incur low overhead: $<5$\% increase for both LUTs and registers.

\section{Using \acro to Enhance \RA \& Related Services}\label{sec:discussion}
We now discuss how \acro can make \RA and related services simpler and more efficient.

\subsection{Constant-Time \RA}\label{sec:constant_RA}
One notable and beneficial feature of \acro is that,
most of the time, \RA no longer needs to be computed over the entire $AR$, 
which significantly reduces \RA execution time on \prv.

If \vrf already knows $AR$ contents from a previous attestation result, 
it suffices to show that $AR$ was not changed since then.
This can be done by attesting $LMT$ \textbf{by itself}, instead of $AR$ in its entirety, 
resulting in substantial reduction of computation time from linear in the size of $AR$ to constant: $|LMT|$, i.e., $32$ bytes.
As such, \RA is performed differently, in two possible cases:

 \noindent-- \textbf{Case-1:} if no modification to $AR$ happened since the last attestation (denoted by $t_{att}$), call  \attest on 
 $LMT$ region only. \vrfy checks for $\utoken \equiv \hmac(KDF(\attkey, \chal), LMT)$. 
 \vrf then learns whether $AR$ was modified since the previous measurement, solely based on $LMT$.
 By checking that $LMT$ corresponds to $t_0 < t_{att}$, this result confirms that $AR$ remained the same in the interim. 
 Therefore, measuring $AR$ again is unnecessary and doing so would be redundant.
 
 \noindent-- \textbf{Case-2:} If $AR$ was modified since the last attestation, call \attest covering entire $AR$. \vrfy is computed 
 normally as described in Constructions~\ref{def:cons_RTC} or~\ref{def:cons_clockless}, depending on the 
 implementation, i.e., \acroa or \acrob.

\noindent\emph{\textbf{Remark 5:} Note that \prv's \RA functionality can easily detect whether $AR$ was modified (in order to decide 
between attesting with \textbf{Case-1} or \textbf{Case-2}) by checking the value of $LMT$, which is readable in software, though not writable.}

Most of the time, \prv is expected to be in a benign state (i.e., no malware), especially if \sadv knows that its presence is 
guaranteed to be detectable. 
In such times, size of attested memory can be reduced 
reduced from several KBytes (e.g., when $AR$ is the entire program memory on a low-end \prv) to a mere $32$ Bytes ($LMT$ size), 
Figure~\ref{fig:rata_savings} depicts an empirical result on the MSP430 MCU showing how this 
optimization can significantly reduce \RA runtime overhead.

In the rest of this section, we discuss some implications of this optimization, along with security improvements offered by 
\acro, to different branches of \RA and related security services.

\subsection{Atomicity \& Real-Time Settings}\label{sec:app1}
Security of hybrid \RA architectures generally depends on {\it temporal consistency} of attested memory. 
Simply put, temporal consistency means ``no modifications to $AR$ during \RA computation''.
Lack thereof allows self-relocating malware to move itself within \dev's memory during attestation, in order to avoid detection, e.g., 
if malware interrupts attestation execution, relocates itself to the part of $AR$ that has already been covered by the
integrity-ensuring function ($HMAC$ in our case), and restarts attestation. 

In higher-end devices, memory locking can be used to prevent modifications until the end of attestation, 
as discussed in~\cite{carpent2018temporal}. However, in low-end devices, where applications run on bare-metal and 
there is no architectural support for memory locking, temporal consistency is attained by enforcing that 
attestation software (\sw) runs atomically: once it starts, it can not be interrupted by any software running on \prv, thus
preventing malware from interrupting \RA and relocating itself.
While effective for security purposes, this requirement conflicts with real-time requirements if \dev serves a
safety-critical and time-sensitive function. 

Some prior  remediation techniques proposed to enable interrupts while maintaining temporal consistency, with high probability. 
SMARM~\cite{smarm} is one such approach.  (Others similar techniques are discussed in~\cite{carpent2018reconciling}). SMARM 
divides attested memory ($AR$) into a set of blocks which are attested in a randomized order. Attestation of one block remains 
atomic. However, interrupts are allowed between attestation of two blocks. Assuming that malware can not guess the 
index of the next block to be attested, even if interrupts are allowed, malware only has a certain probability of avoiding detection. 
If the entire attestation procedure is repeated multiple times, this probability can be made arbitrarily small. 

We note that, given the \acro optimization discussed in Section~\ref{sec:constant_RA}, attestation can be computed faster. 
In particular, since most Pseudo Random Function (PRF) implementations use block sizes of at least 32 bytes, the atomic attestation 
of one block in a SMARM-type strategy can not be faster than the attestation on $LMT$ in \acro ($|LMT| = 32$ Bytes). 
In addition, attestation of $LMT$ provides information about the content of $AR$ in its entirety, with no probability of evasion. 
We believe this makes \acro more friendly to safety-critical operations than existing approaches.

In such settings, we envision that $AR$ would be attested in its entirety at system boot time (\textbf{Case-2} in 
Section~\ref{sec:constant_RA}), while subsequent \RA would be computed on $LMT$ only (\textbf{Case-1} in 
Section~\ref{sec:constant_RA}). We note that, if $AR$ is eventually modified, \prv would need to fall back to  
\textbf{Case-2} for the next \RA computation, which takes time to run atomically. However, after an unauthorized 
modification to \prv's memory, it is unclear why one would still want to offer real-time guarantees to compromised software. 

\vspace{-2mm}
\subsection{Collective \RA Protocols and Device-to-Device Malware Relocation}\label{sec:app3}
Collective \RA protocols (CRA) (aka swarm attestation)~\cite{SEDA,SANA,LISA,DARPA,SCAPI,SALAD,SAP} are a set of 
techniques that attest a large number of devices that operate together as a part of a larger system. CRA schemes typically 
assume hybrid \RA architectures on individual devices and look into how to attest many devices efficiently.
One security problem that is typically out of scope on single-device \RA and becomes relevant in CRA settings is caused 
by migratory malware. This is an analog of intra-device self-relocating malware (discussed in Section~\ref{sec:app1}) that 
appears in collective settings. Specifically, instead of moving around inside the memory of the same device, it migrates 
from device to device to avoid detection.

To guarantee detection of migratory malware, 
CRA result must convince \vrf that all devices were in a safe state \textbf{within the same time window}, implying that 
malware had no destination device to which to migrate and avoid detection. Consequently, if a single-device attestation 
result conveys a safe state only at some point in between the execution of \request and \vrfy algorithms, 
it is nearly impossible (especially, in the presence of network delays) to conclude that migratory malware is not present in the 
swarm. Although this problem is discussed in the CRA literature existing approaches either place it outside their adversarial model~\cite{LISA,SEDA,SANA,SALAD}, 
or make a strong assumption about clock synchronization among all devices in the swarm~\cite{ibrahim2017seed,DARPA,SCAPI,SAP},
so that all devices can be scheduled to run \attest at the same time.

\vspace{2mm}
\begin{mdframed}
\vspace{-1em}
	\begin{construction}[CRA-\acro]\label{def:cons_CRA_rata}
	\footnotesize
		Let $S = \{\prv_1,...,\prv_n\}$ denote a swarm of $n$ devices individually equipped with \acrob hybrid \RA facilities. 
		Let $LMT_i$ be the value of LMT in $\prv_i$. Also, $\vrfy(\prv_i)$ denotes the verification algorithm of 
		Construction~\ref{def:cons_clockless} for $\prv_i$.
		Consider a protocol in which:
		\begin{enumerate}
		 	\item \vrf executes \acrob protocol, as defined in Construction~\ref{def:cons_clockless} with each $\prv_i$ in parallel. 
			Let $t(Req_i)$ denote the time when \vrf issued the request to $\prv_i$.
		 	\item \vrf collects all responses and computes $\vrfy(\prv_i)$ for all $\prv_i \in S$. It then uses the values of $LMT_i$ to 
			learn ``since when'' $\prv_i$ has been in a valid state. We denote this time as $t(LMT_i)$.
		\end{enumerate}
	\end{construction}
\end{mdframed}
\vspace{2mm}

We argue that, by addressing the \toctou problem in the single-device setting, \acrob can be utilized to construct the first CRA protocol secure against migratory malware without relying on synchronization of the entire swarm. 
To see why this is the case, consider Construction~\ref{def:cons_CRA_rata}. In this construction, \toctou-Security on individual devices allows \vrf to conclude that each \dev was in a valid state within a fixed time interval. Therefore, by checking 
the overlap in the valid interval of all \dev-s, \vrf can learn the time window in which the entire swarm was safe as a whole, or 
detect migratory malware when such time window does not exist. Theorem~\ref{th:rata_CRA_sec} states the concrete guarantee 
offered by Construction~\ref{def:cons_CRA_rata}.

\vspace{2mm}
\begin{mdframed}
\vspace{-1em}
\textit{\begin{theorem}\label{th:rata_CRA_sec}
\small
In Construction~\ref{def:cons_CRA_rata}, if for all $\prv_i \in S$, $\vrfy(\prv_i)$ in step 2 succeeds for some $t(LMT_i)$, then it must be the case that entire $S$ was in a valid state in the time window defined by the interval:
\begin{equation}\label{eq:interval}
\footnotesize
 (\textbf{max}[t(LMT_1), ...,t(LMT_n)]~,~\textbf{min}[t(Req_1), ...,t(Req_n)])
\end{equation}
assuming equation~\ref{eq:interval} constitutes a valid interval.\\
{\bf Note:} $(a,b)$ is a valid interval if $a < b$.
\end{theorem}}
\end{mdframed}
\vspace{-1mm}
\begin{proof}\textit{(Sketch)}~It follows directly from the observations that:

  \noindent -- Given \RA-Security, for each $\prv_i \in S$, a valid response can not be produced before the time when $\prv_i$ 
  receives \chal, which is strictly greater than $t(Req_i)$.
  
  \noindent -- Given \toctou-Security, for each $\prv_i \in S$ with  $\vrfy(\prv_i) = 1$, its memory could not have been changed 
  between $t(LMT_i)$ and the first call to \attest after $t(Req_i)$.
\end{proof}

\subsection{Runtime Attestation}\label{sec:app4}
Runtime attestation focuses on detection of runtime/data-memory attacks, providing authenticated information about software 
execution on \prv. While it seems unrelated to detection of retrospective program memory modifications, we argue
that \acro can also offer improvement to runtime attestation techniques.

Proofs of execution (PoX) for embedded systems were recently explored in~\cite{apex} (APEX). They are used to prove that a given operation on \prv was performed through the
execution of the expected code and to verify that outputs were indeed produced by this execution. Control Flow Attestation (CFA) introduced in~\cite{cflat} (C-FLAT) allows \vrf to also verify whether software that executed on \prv took a specific (or a set of) valid control path(s) enabling detection of ROP/code-reuse type attacks to vulnerable code.

We note that regular (or static) \RA is a common stepping stone in these respective functionalities. In C-FLAT, OAT~\cite{sun2020oat}, and Tiny-CFA~\cite{tinycfa}, the executable must be instrumented with specific instructions  to enable CFA and \RA is used to verify 
that such instructions were not removed or modified. Besides, even executions with the same control-flow may differ in 
behavior/outputs if their instructions differ.
Similarly, in APEX execution is proven to \vrf with attestation of execution metadata.
However, without attesting the corresponding executable (in program memory), this proof would have no meaning other than: ``some code executed successfully''.

In many applications, the same executable is expected to remain in memory for long periods of time, 
while its proper execution (or control-flow) must be verified repeatedly, per safety-critical embedded operation~\cite{sun2020oat}. 
\acro's optimization discussed in~\ref{sec:constant_RA} can minimize the overhead of such successive runtime attestations.

To illustrate this concept we combined \acro with APEX and Tiny-CFA (which itself is implemented atop APEX). In APEX, all runtime overhead \textit{vis-a-vis} cost of executing the same software without proving its successful execution to \vrf is caused by the cost of static \RA.
Since APEX is implemented atop \vrased, we were able to implement a \acro-compliant version of APEX without changing the internal behavior of \acro's hardware modules nor APEX hardware module itself. As such, this approach substantially reduced PoX and CFA computational costs (these savings are consistent with Figure~\ref{fig:rata_savings}) while requiring the same additional hardware cost as reported in Table~\ref{tab:exp}.

\section{Related Work}\label{sec:rw}
\noindent -- \textbf{Remote Attestation (\RA):} 
\RA techniques generally fall into three categories: hardware-based, software-based and hybrid. Hardware-based techniques~\cite{PFM+04, tpm, KKW+12, NAD+13}
either perform \RA using a dedicated autonomous hardware component (e.g., a TPM~\cite{tpm}),  or require substantial
changes to the underlying instruction set architecture in order to support execution of trusted software (e.g., SGX~\cite{sgx}).
Such changes are too expensive for cost-sensitive low-end embedded devices. On the other end of the spectrum,  software-based 
techniques~\cite{KeJa03, SPD+04, SLS+05} require no hardware security features; they perform \RA using a custom checksum function implemented entirely in software. 
Security of software-based techniques relies on a precise measurement timings, which is only applicable to settings where 
the communication delay between \vrf and \dev is negligible and/or constant, e.g., communication between peripherals and a 
host CPU. Thus, software-based \RA is unsuitable for environments where \RA must be performed over the internet. 
Whereas, hybrid \RA is particularly suitable for low-end embedded devices. It provides the same security guarantees as 
hardware-based \RA, while minimizing modifications to underlying MCU hardware. 
Current hybrid \RA techniques~\cite{smart,trustlite,hydra,vrasedp,FNR+14,tytan} implement the 
integrity-ensuring function (e.g., MAC) in software, and
use trusted hardware to control execution of this software, preventing any violations that might cause \RA security problems, e.g.,
gadget-based attacks~\cite{rop} or key leakage. This paper represents a paradigm shift of hybrid \RA, by having trusted 
hardware additionally providing some context about \prv's memory state.\\
\noindent -- \textbf{Temporal Aspects of \RA:} 
Besides \toctou, two other temporal aspects are essential for \RA security: First, temporal 
consistency~\cite{carpent2018temporal} means guaranteeing that the \RA result reflects an instantaneous 
snapshot of \prv's attested memory at some point in time during \RA. Lack thereof allows self-relocating malware 
to escape detection by copying and/or erasing itself during \RA.
Temporal consistency is achieved by enforcing atomic (uninterruptible) execution of attestation code, or by 
locking attested memory (i.e., making it unmodifiable) during \RA execution. Second, when \RA is used on 
safety-critical and/or real-time devices~\cite{carpent2018reconciling}, atomicity requirement might interfere 
with the real-time nature of \prv's application. To address this issues, SMARM~\cite{smarm} relaxes this requirement 
by using probabilistic malware detection. Meanwhile, ERASMUS~\cite{erasmus} and SeED~\cite{ibrahim2017seed} 
are based on \prv's self-measurements, in order to detect transient malware that infects \prv and leaves before the 
next \RA instance. See Section~\ref{sec:motivation} for further discussion on these types of techniques. 
Atrium~\cite{zeitouni2017atrium} deals with physical-hardware adversaries that intercept instructions as they 
are fetched to the CPU during attestation. Atrium refers to that issue as \toctou. Despite nomenclature, 
that issue is clearly different from \acro's goal.\\
\noindent -- \textbf{Formal Verification and \RA:} 
Formal verification provides significantly higher level of assurance, yielding provable security for protocol 
specifications and implementations thereof.
Recently, several efforts focused on formal verification of security-critical services and systems
~\cite{hawblitzel2014ironclad,beringer2015verified,hacl,bhargavan2013implementing,compcert,sel4}.
\vrased~\cite{vrasedp} realized a formally verified \RA architecture targeting low-end devices.
Other formally verified security services were obtained by extending \vrased to derive remote proofs of software update, 
memory erasure and system-wide MCU reset~\cite{pure}. APEX~\cite{apex} builds on top of \vrased to 
develop a verified architecture for proofs of remote software execution on low-end devices~\cite{apex}.
RATA also builds on top of \vrased, extending it to provide \toctou security while retaining original verified guarantees. 
Relying on \vrased allows us to reason about \acro design and to 
formally verify its security properties.
Nonetheless, \acro's main concepts are applicable to other hybrid (and possibly hardware-based, such as~\cite{Sancus17}) \RA architectures.

\section{Conclusions}\label{sec:conclusion}
In this paper, we design, prove security of, and formally verify two designs (\acroa~and \acrob) to secure \RA against \toctou-related attacks, which perform illegal binary modifications on a low-end embedded system, in between attestation measurements.
\acroa and \acrob modules are formally specified and verified using a model-checker.
They are also composed with \vrased~-- a verified \RA architecture.
We show that this composition is \toctou-secure using a reduction-based cryptographic proof.
Our evaluation demonstrates that a \toctou-Secure design is affordable even for cost-sensitive low-end embedded devices. Additionally, in most cases, it reduces \RA time complexity from linear to constant, in the size of the attested memory.


\balance
\bibliographystyle{ieeetr}
\bibliography{references}
\appendix

\vspace{3mm}
\noindent\textbf{\Large{APPENDIX}}

\section{\vrf authentication details}\label{apdx:vrased_auth}

\lstset{language=C,
	basicstyle={\footnotesize\ttfamily},
	showstringspaces=false,
	frame=single,
	xleftmargin=2em,
	framexleftmargin=3em,
	numbers=left, 
	numberstyle=\tiny,
	commentstyle={\tiny\itshape},
	keywordstyle={\tiny\bfseries},
	keywordstyle=\color{blue}\tiny\ttfamily,
	stringstyle=\color{red}\tiny\ttfamily,
        commentstyle=\color{black}\tiny\ttfamily,
        morecomment=[l][\color{magenta}]{\#},
        breaklines=true
}
\begin{figure}[!hbtp]
\begin{lstlisting}[basicstyle=\tiny, numberstyle=\tiny]
void Hacl_HMAC_SHA2_256_hmac_entry() {
    uint8_t key[64] = {0};
    uint8_t verification[32] = {0};
    if (memcmp(CHALL_ADDR, CTR_ADDR, 32) > 0)
    {
	memcpy(key, KEY_ADDR, 64);
	
	hacl_hmac((uint8_t*) verification, (uint8_t*) key,
		  (uint32_t) 64, *((uint8_t*)CHALL_ADDR) ,
		  (uint32_t) 32);
		  
	if (!memcmp(VRF_AUTH, verification, 32)
	{
	    hacl_hmac((uint8_t*) key, (uint8_t*) key, 
	    	(uint32_t) 64, (uint8_t*) verification, 
	    	(uint32_t) 32);
	    hacl_hmac((uint8_t*) MAC_ADDR, (uint8_t*) key, 
	    	(uint32_t) 32, (uint8_t*) ATTEST_DATA_ADDR, 
	    	(uint32_t) ATTEST_SIZE);
	    memcpy(CTR_ADDR, CHALL_ADDR, 32);
	}
    }

    return();
}
\end{lstlisting}
\caption{\sw Implementation with \vrf authentication~\cite{vrasedp}.}\label{fig:sw_att_code_auth}
\end{figure}

To prevent an adversary from impersonating \vrf and sending fake attestation requests to \dev, \vrased design supports authentication of \vrf as part of \sw execution. The implementation is based on the protocol in~\cite{brasser2016remote}. In this protocol, \chal is chosen by \vrf as a monotonically increasing nonce. As such, for subsequent requests $i$ and $i+1$, it is always the case that $\chal_i < \chal_{i+1}$.

Figure~\ref{fig:sw_att_code_auth} shows \vrased \texttt{C} implementation of \sw, including \vrf authentication.
It also builds upon HACL* verified HMAC to authenticate \vrf, in addition to computing the authenticated integrity check over $AR$. 
In this case, \vrf's request additionally contains an \hmac of the challenge computed using \attkey.
Before calling \sw, software running on \dev is expected to store the received challenge on a fixed address $CHALL\_ADDR$ and the corresponding received HMAC on $VRF\_AUTH$.
\sw discards the attestation request if (1) the received challenge is less than or equal to the latest challenge, or (2) HMAC of the received challenge is mismatched.
After that, it derives a new unique key using HKDF~\cite{krawczyk2010hmac} from \attkey and the received \hmac
and uses it as the attestation key. 

To support secure authentication, \vrased extends \hw with two additional properties to make the memory region that stores \dev's counter immutable to untrusted applications (any software except \sw).
Notably, the counter requires persistent and writable storage, because \sw needs to modify it at the end of each attestation execution.

\section{Proof of Theorem~\ref{th:toctou_sec}}\label{apdx:proof_rataa}

\begin{proof}
By contradiction, assume a polynomial \sadv that wins the game in Definition~\ref{def:toctou_sec} with 
probability $Pr[\adv, \text{\RA-\toctou-game}] > \negl[l]$. Therefore, \sadv can produce $t_{LMT}||\utoken_{\sadv}$ such that:
	\begin{align*}
	\vrfy^{\vrf}\OParan \utoken_{\sadv}, \chal, M, t_0, t_{LMT} \CParan = 1 \\ \text{ and }\\ \exists_{t_0 \leq t_i \leq t_{att}}\{AR(t_i) \neq  M \}	 
	\end{align*}
By definition, \vrfy in Construction~\ref{def:cons_RTC} results in $1$ only if $t_{LMT} < t_0$.
If \sadv simply replies with the actual value $t_{LMT} = LMT \geq t_i$, \vrfy result would be $0$,
since $t_i \geq t_0$, failing to satisfy \vrfy condition: $t_{LMT}<t_0$.
Thus, to obtain $\vrfy=1$, \sadv must spoof the value of $t_{LMT}$ to $t_{LMT} < t_0$.

Upon receiving the spoofed value of $t_{LMT}$ the \vrfy now expects:
\begin{align}
\utoken_{\adv} \equiv \hmac(KDF(\attkey, MR), M)
\end{align}
where expected $M$ reflects $LMT=t_{LMT}$, i.e., $LMT < t_0$.

Also, hardware enforced properties~\ref{eq:prop1} and~\ref{eq:prop2} guarantee that $LMT \in AR$ 
always contains the time of the most recent modification of $AR$. Thus, because $t_{att} \geq t_i$, it 
must be the case that $AR(t_{att})$ reflects $LMT \geq t_i$ implying $LMT \neq t_{LMT}$, and 
consequently $AR(t_{att}) \neq M$.

Under such restriction, \adv ability to win the game implies its capability to produce $\utoken_{\adv}$ such 
that $\vrfy^{\vrf}\OParan \utoken, \chal, M, t_0, t_{LMT} \CParan = 1$, even though modifying $AR$ 
such that $AR(t_{att}) = M$ is not possible. To conclude the proof, we show that the existence of 
such an \sadv implies the existence of another adversary $\adv_{\RA}$ that wins the \RA security 
game in Definition~\ref{def:vrased_sec} against \vrased, contradicting the theorem's assumption.

To win the game in Definition~\ref{def:vrased_sec} $\adv_{\RA}$ behaves as follows:
\begin{enumerate}
 \item At time $t_i$ where $t_0 \leq t_i \leq t_{att}$, $\adv_{\RA}$ modifies $AR$ causing $LMT \in AR$ 
 to store the value of $t_i$.
 \item $\adv_{\RA}$ receives \chal from the challenger in step (2) of \RA security game of Definition~\ref{def:vrased_sec} 
 and executes the same algorithm of \sadv with inputs \chal and $t_{att}=t$ to produce $\utoken_{\sadv}$, 
 such that $\vrfy^{\vrf}\OParan \utoken_{\sadv}, \chal, M, t_0, t_{LMT} \CParan = 1$ with probability:
 \begin{center}
 $Pr[\adv, \text{\RA-\toctou-game}] > \negl[l]$,
 \end{center}
 even though $t_{LMT} < t_0 < t_i$.
 \item As a response in step 3 of the game in Definition~\ref{def:vrased_sec}, $\adv_{\RA}$ replies with:
  $\sigma = \utoken_{\adv}$.
\end{enumerate}
Since $\vrfy^{\vrf}\OParan \utoken_{\sadv}, \chal, M, t_0, t_{LMT} \CParan = 1$, it follows that $\sigma = \utoken_{\adv} = \hmac(KDF(\attkey, MR), M)$, for expected $M$ containing $LMT=t_{LMT}$.
However, due to the $AR$ modification at time $t_i$, $AR(t)$ must reflect $LMT \geq t_i$, satisfying the condition that $AR(t) \neq M$ and allowing $\adv_\RA$ to win the game in Definition~\ref{def:vrased_sec} with probability:
\begin{equation}
 Pr[\adv,\text{\RA-game}] = Pr[\adv,\text{\RA-\toctou-game}]>\negl[l]
\end{equation}
\end{proof}

\section{Proof of Theorem~\ref{th:toctou_sec_b}}\label{apdx:proof_ratab}

We show that, if properties in Equations~\ref{eq:2prop1}, \ref{eq:2prop2} and~\ref{eq:2prop3} hold, existence of \adv that wins the \toctou 
security game against \acrob implies the existence of another \adv that wins \RA security game against \vrased, thus 
contradicting the initial premise.

\begin{proof}
	By contradiction, assume a polynomial \sadv that wins the game in Definition~\ref{def:toctou_sec} with 
	probability $Pr[\adv, \text{\RA-\toctou-game}] > \negl[l]$. Therefore, \sadv is able to produce response $LMT_{\sadv}||\utoken_{\sadv}$ such that:
	
	\begin{align*}
	\vrfy^{\vrf}\OParan \utoken_{\sadv}, \chal, M, t_0, T , LMT_{\sadv}\CParan = 1 \\ \text{ and }\\ \exists_{t_0 \leq t_i \leq t_{att}}\{AR(t_i) \neq  M \}	 
	\end{align*}
	
	By definition, in Construction~\ref{def:cons_clockless}, \vrfy outputs $0$ if $LMT_\sadv$ differs from $\chal_P$ stored by \vrf in the challenge-time association pair $P=(\chal_P, t_{P})$.
	If $LMT_\sadv = \chal_P$, it corresponds to a challenge value sent before $t_0$ (assuming sensible choices of $t_0$ by \vrf ).
	Therefore, in order to win, \adv \textbf{must choose $LMT_\sadv = \chal_P$}.
	
	Since $LMT \in AR$, by claiming a value for $LMT_{\adv}$ fitting the restriction above, \adv causes the expected memory value $M$ to also reflect, $LMT=LMT_{\adv}$.
	At this point, \adv has two possible actions: to modify $AR$ to call \attest with $AR(t_{att})=M$; or to obtain $\utoken_{\sadv}$ even with $AR(t_{att})\neq M$. First we show that the latter is \adv's only option.
	
	Let us say that \adv attempts to set $AR(t_{att})=M$ to call \attest.
	In this case, we highlight three observations about \acrob:
    \begin{compactenum}
     \item By LTL statement~\ref{eq:2prop3}, any modification to $AR$ in between the $i$-th and $(i+1)$-th authenticated computations of \attest, will cause $AR$ to change to reflect $LMT=\chal_{i+1}$ in following \RA responses. Therefore, the premise that
    \begin{align*}
    \exists_{t_0 \leq t_i \leq t_{att}}\{AR(t_i) \neq  M \}	 
	\end{align*}
     will necessarily update $LMT$.
     \item From \vrased authentication (see Appendix~\ref{apdx:vrased_auth}), for subsequent \RA challenges $\chal_i$ and $\chal_{i+1}$ that authenticate successfully, it is always the case that $\chal_i < \chal_{i+1}$.
     \item From LTL statement~\ref{eq:2prop2}, \acrob never updates $LMT$ with a challenge if it does not authenticate successfully. Since authentication implies $\chal_i < \chal_{i+1}$, a call to \attest never causes LMT to be updated to a previously used \chal.
    \end{compactenum}
    
    From observations 1, 2, and 3 above, it is impossible to set $AR=M$ by calling \attest, because any modification to LMT caused by \attest will always change $LMT$ to a value that was never used before and thus different from $\chal_P$. At this point \adv last resource is to try to write to $LMT$ directly. However, this is immediately in conflict with LTL property~\ref{eq:2prop1}.

	Since making $AR(t_{att})=M$ is impossible after a modification at time $t_i$, the assumption that \adv wins the game in Definition~\ref{def:toctou_sec} implies that \adv is able to produce $\utoken_{\sadv}$ that verifies successfully even when $AR(t_{att}) \neq M$.
	To conclude the proof, we show that existence of such \sadv implies existence of another adversary $\adv_{\RA}$ that wins the \RA security game in Definition~\ref{def:vrased_sec}.
	
	To win the game in Definition~\ref{def:vrased_sec} $\adv_{\RA}$ is constructed as follows:
	\begin{enumerate}
		\item At time some $t_i$, where $t_0 \leq t_i \leq t$, $\adv_{\RA}$ modifies memory in $AR$.
		\item $\adv_{\RA}$ receives $\chal$ in step 2 of \RA security game of Definition~\ref{def:vrased_sec}, and executes the same algorithm as \sadv on \chal and with $t_{att} = t$ to produce $\utoken_{\sadv}$ such that $\vrfy^{\vrf}\OParan \utoken_{\sadv}, \chal, M, t_0, T , LMT_{\sadv}\CParan = 1$ 
		with probability:
		\begin{center}
			$Pr[\adv, \text{\RA-\toctou-game}] > \negl[l]$.
		\end{center}
		\item As a response in step 3 of the game in Definition~\ref{def:vrased_sec}, $\adv_{\RA}$ replies with $\sigma = \utoken_{\adv}$.
		%
	\end{enumerate}
	Since $\vrfy^{\vrf}\OParan \utoken_{\sadv}, \chal, M, t_0, T , LMT_{\sadv}\CParan = 1$, it follows that $\sigma = \hmac(KDF(\attkey, \chal), M)$ (first condition for $\adv_{\RA}$ to win), for expected $M$ containing $LMT=LMT_{\sadv}$.
	On the other hand, because memory was modified at time $t_i$, it must be the case that $AR(t)$ has $LMT \neq LMT_{\sadv}$. Thus satisfying the remaining condition that $AR(t)\neq M$ implying that $\adv_{\RA}$ wins the game in Definition~\ref{def:vrased_sec} with probability:
	\begin{equation}
	Pr[\adv,\text{\RA-game}] = Pr[\adv,\text{\RA-\toctou-game}]>\negl[l]
	\end{equation}
\end{proof}

\section{\acro Implementation with SANCUS}\label{apdx:sancus}
To demonstrate \acro generality, we also implemented it atop SANCUS~\cite{Sancus17}: a hardware-based \RA 
architecture targeting the same class of embedded devices. To the best of our knowledge, aside from VRASED 
(used in our verified implementation), SANCUS is the only other open-source \RA architecture for low-end embedded 
systems, which justifies our choice. We note that this implementation is intended to demonstrate \acro generality 
and that provable security guarantees derived from \acro-with-\vrased do not apply here. Since SANCUS does 
not provide a formal security model and analysis, provable composition of RATA atop SANCUS is not currently possible. 

Since \acro operates as a standalone monitor that does not interfere with neither the CPU nor the underlying 
\RA architecture functionality, adapting RATA to work with SANCUS is almost effortless.
We describe this implementation in terms of \acroa, which is simpler and does not depend on \vrf's authentication.
The main difference from the \vrased-based implementation is due to SANCUS support for isolated software 
modules (SMs), where each SM is attested individually as an independent program.
We note that even SANCUS' support for attestation and inter-process isolation is insufficient to provide \toctou-Security, 
since \prv's program memory could be physically re-programmed or modified via exploits to vulnerabilities in the 
code of the isolated application itself, without \vrf's knowledge.
Hence, similar to \vrased's \RA case,  \acro also complements SANCUS security guarantees.

To enable \acro functionality over SANCUS one must be careful (when programming \prv) to configure the 
software binary such that the program memory of a particular SM of interest coincides with \acro's $AR$ region. 
As such, program memory of the SM will be automatically checked by \acro module and SANCUS attestation 
of such SM's program memory will also cover $LMT$ (since $LMT \in AR$) providing an authenticated proof to 
\vrf of the time of the latest modification of such SM's program memory.

We note that this approach requires one \acro module per SM, since multiple SMs imply dividing \prv's 
program memory into multiple $ARs$ and corresponding LMT regions. Nonetheless, since low-end devices 
typically run very few processes, we expect the cost to remain manageable.

Because SANCUS is implemented on the same MCU as \vrased (OpenMSP430), no internal modifications 
are required to \acro hardware module, and its additional hardware cost remains consistent with that 
reported in Table~\ref{tab:exp}. To support \toctou-Secure attestation of multiple SMs, this cost grows linearly, 
i.e., the cost incurred by one \acro hardware module multiplied by the number of independent SMs that 
should support \toctou-Secure attestation. We note that, in \acroa's case, the same secure read-only 
synchronized clock can be shared by all such modules.

\end{document}